\documentclass[preprint]{aa}
\usepackage{graphicx}        % For eps figures, newer & more powerfull
\usepackage{natbib}         % For citations: redefine \cite commands
\usepackage{amssymb}        % useful mathemati cal symbols
\usepackage{amsmath}
\usepackage[scanall]{psfrag}%\usepackage{color}           % For color text: \color command
\usepackage{threeparttable}
\usepackage{xcolor}
\usepackage[normalem]{ulem}
\usepackage{MR_AA}
\usepackage{interval}
\usepackage{soul}

\providecommand\bnabla{\boldsymbol{\nabla}}

\newcommand\bg{{\boldsymbol{g}}}

\newcommand\bF{{\boldsymbol{F}}}

\newcommand\om{$\omega$-model}

\def\Tau{{\rm T}}

\newcommand{\Tej}{T_{\rm eff}^{\rm jump}}
\newcommand{\Teff}{T_{\rm eff}}
\newcommand{\geff}{g_{\rm eff}}

\begin{document}

        \titlerunning{Critical angular velocity and anisotropic mass loss of rotating stars}
    \title{Critical angular velocity and anisotropic mass loss of rotating stars with radiation-driven winds}
   % \subtitle{I. Critical angular velocity at the $\Omega\Gamma$-limit and effects of rotation on radiation-driven mass loss at ZAMS}
\author{D. Gagnier\inst{1} \and M. Rieutord\inst{1} \and
C. Charbonnel\inst{2,1} \and B. Putigny\inst{1} \and F. Espinosa
Lara\inst{3}}

\institute{
IRAP, Universit\'e de Toulouse, CNRS, UPS, CNES,
14, avenue \'{E}douard Belin, F-31400 Toulouse, France
\and
Department of Astronomy, University of Geneva, Chemin des Maillettes 51,
1290, Versoix, Switzerland 
\and
Space Research Group, University of Alcal\'a, 28871 Alcal\'a de Henares, Spain
\\
\email{[Damien.Gagnier,Michel.Rieutord]@irap.omp.eu, 
Corinne.Charbonnel@unige.ch}
    }

\abstract
{The understanding of the evolution of early-type stars is tightly related to that of the effects of rapid rotation. For massive stars, rapid rotation combines with their strong radiation-driven wind.}{The aim of this paper is to investigate two questions that are prerequisite to the study of the evolution of massive rapidly rotating stars: (i) What is the critical angular velocity of a star when radiative acceleration is significant in its atmosphere? (ii) How do mass and angular momentum loss depend on the rotation rate?}
{To investigate fast rotation, which makes stars oblate, we used the 2D ESTER models and a simplified approach, the \om, which gives the latitudinal dependence of the radiative flux in a centrifugally flattened radiative envelope.}
{We find that radiative acceleration only mildly influences the critical angular velocity, at least for stars with masses lower than $40~\msun$. For instance, a 15~$M_\odot$ star on the zero-age main sequence (ZAMS) would reach criticality at a rotation rate equal to 0.997 the Keplerian equatorial rotation rate. We explain this mild reduction of the critical angular velocity compared to the classical Keplerian angular velocity by the combined  effects of gravity darkening and a reduced equatorial opacity that is due to the centrifugal acceleration. To answer the second question, we first devised a model of the local surface mass flux, which we calibrated with previously developed 1D models. The discontinuity  (the so-called bi-stability jump) included in the $\dot{M}-\teff$ relation of 1D models means that the mass flux of a fast-rotating star is controlled by either a single wind or a two-wind regime. Mass and angular momentum losses are strong around the equator if the star is in the two-wind regime. We also show that the difficulty of selecting massive stars that are viewed pole-on makes detecting the discontinuity in the relation between mass loss and effective temperature also quite challenging.}{} 


\maketitle

\section{Introduction}

Among the numerous problems that need to be overcome when stars are modelled, those
related to rotation are of particular nature in the frame of classical
1D models because rotation breaks the imposed spherical symmetry. Rotating stars are indeed not only distorted by the centrifugal acceleration,
but are also pervaded by large-scale flows that carry chemical
elements and angular momentum. The importance of these effects has been
appreciated for quite some time now \cite[e.g.][and references therein]{MaederMeynet2000}, and  specific
modelling simplifications are usually included in 1D stellar evolution codes to reproduce the
expected effects of global rotation. 
For instance, the transport of angular momentum that results from small-scale turbulence and
large-scale circulation induced by rotation in radiative zones is
inserted in 1D evolution models either as an advection-diffusion process following \cite{Zahn1992}, \cite{MeynetMaeder1997}, and \cite{MZ98} (e.g.
Geneva code, \citealt{eggenberger2008};  STAREVOL, \citealt{Decressin_etal2009}, \citealt{Amard_etal16}; FRANEC, \citealt{ChieffiLimongi13}; CESTAM, \citealt{marques_etal13}) or as a purely
diffusive process (e.g. {\it{Kepler}}, \citealt{Hegeretal00Keplercode}; STERN, \citealt{YoonLanger05}; MESA, \citealt{paxton_etal13}). % ; STAREVOL, \citealt{Amard_etal16};). 
The associated transport of chemicals (so-called rotation-induced mixing) is always treated as a diffusive process \citep[as justified by][]{chaboyer_Z92}. 

This  modelling of rotation effects is only justified
for slow rotators \cite[][]{zahn92}. Early-type stars are, however, often considered to be fast rotators. The hypotheses and approximations of current prescriptions are therefore no longer valid for such stars. Be-type stars, for instance, are
well known to be fast rotators close to the break-up limit \cite[e.g.][]{PR03,Bastian2017}, that is, the centrifugal force nearly balances gravity at equator.
These stars show evidence of mass loss that is associated with their near
break-up rotation \cite[][]{carciofi_etla08,Krticka2011, Riviniusetal13,Georgy2013, Granada2013,Granadaetal14}. Furthermore, early-type stars may also be very luminous and therefore have high radiation pressure at their surface. The induced radiation-driven wind is responsible for a significant loss of mass and angular momentum, which notably influences the evolutionary paths of massive stars \citep[e.g.][]{Langer1998, Vink2010}.
Because of gravity darkening \citep[e.g.][]{ELR11}, mass loss from rotating massive stars is expected to be anisotropic \citep[e.g.][]{Owocki1996, Owocki1997, Pelupessy2000, MaederMeynet2000, Georgy2011}. These anisotropies in  turn affect the evolution of rotation and are likely to play a significant role in the interior dynamics of massive stars \citep[][]{zahn92, Maeder1999, Lignieres2000,Lau_etal11, RB14}.

%{\sl{Studying fast rotating massive stars with 1D models is thus hazardous regarding the approximations used.}} 
The treatment of fast rotation thus requires developments beyond the current model approximations, although this approach has been extremely useful to make significant progress in the field \citep[e.g.][and references therein]{MM2015IAUS}. In this context,
the achievement of the first self-consistent 2D models of rapidly rotating early-type stars, worked out by
Espinosa Lara and Rieutord \cite[e.g.][]{ELR13, Rieutordal2016}, opens the door to exploring the evolution of fast stellar rotators. Such models are expected to provide new constraints on the internal rotation-induced mechanisms as well as on the radiative and mechanical mass loss \citep[e.g.][]{Krticka2011}, which all significantly affect the different predictions and outputs of the 1D stellar evolution models \cite[e.g.][]{Meynet2000,MaederMeynet2000,MaederMeynet2010,Smith14reviewmassloss,Meynetetal15, Renzo2017}.

The present work aims at investigating two questions that are prerequisite to the study of the evolution of massive rapidly rotating stars: (i) What is the critical angular velocity of a star when radiative acceleration is significant in its atmosphere? (ii) How do mass and angular momentum loss in massive stars depend on rotation?

\medskip
This paper is organised as follows. In Sect.~\ref{sec:crit} we reconsider the question of the critical angular velocity in light of ESTER 2D models and the simplified $\omega$-model of \cite{ELR11}.
We then revisit the prescription of mass loss in fast-rotating stars. To this end, we first focus on deriving a local mass-flux prescription based on the 1D CAK \citep{CAK1975} and mCAK \citep{PPK} theories for non-rotating stars (Sect. 3). Next, we generalise this prescription to rotating stars. We compute the latitudinal variations of  mass and angular momentum fluxes with ESTER 2D models and discuss the effects of rotation on global losses of mass and angular momentum (Sect.~\ref{sec:lat}). We finally summarise our answers to the questions that motivated this work (Sect.~\ref{sec:conclu}).

\section{Critical angular velocity and the $\Omega\Gamma$-limit}\label{sec:crit}

\subsection{The $\Omega\Gamma$-limit question}

\subsubsection{Some context}

In stars more massive than $\sim 7~M_\odot$ that are close to solar metallicity, radiative acceleration
plays a significant role in the (assumed) hydrostatic equilibrium.  Total gravity is usually introduced,

\begin{equation}\label{eq:gtot}
\bg_{\rm tot} = \bg_{\rm eff} + \bg_{\rm rad} 
,\end{equation}
where effective gravity (or gravito--centrifugal acceleration) $\bg_{\rm eff}$ is supplemented by
radiative acceleration, 

\begin{equation}\label{eq:grad}
\bg_{\rm rad}= \frac{\kappa \bF}{c}
,\end{equation}
where $\kappa$ is the flux-weighted opacity\footnote{Strictly speaking, $\kappa$ is the mass absorption coefficient, $\kappa=\mu/\rho$, with opacity $\mu=\lambda^{-1}$, where $\lambda$ is the mean free path of photons, and density $\rho$. However, in the following, we use, as is frequently done in the current context, the term ``opacity''.}, which we approximate with the total Rosseland mean opacity, $\bF$ the radiative flux, and $c$ the speed of light.

The so-called $\Omega\Gamma$-limit introduced by \cite{MaederMeynet2000} is
reached when $\bg_{\rm tot}=\boldsymbol{0}$ somewhere on the stellar
surface. It is associated with an actual critical angular velocity
$\Omega_c$ that is different from the Keplerian angular velocity
\begin{equation}
\Omega_k=\sqrt[]{\frac{GM}{R_{eq}^3}}
,\end{equation}
which is the break-up limit (or the $\Omega$-limit) when radiative acceleration can be neglected in total gravity. In this equation, $G$ is the gravitation constant, $M$ is the stellar mass, and $R_{eq}$ is its equatorial radius.

The expression
of the correct critical angular velocity when the effects of radiation
cannot be neglected has been debated lively. For instance, \cite{Langer1997,Langer1998} suggested that stars close to the Eddington limit have a lower critical angular velocity, while \cite{Glatzel1998} stressed that the Eddington parameter, namely%\LEt{please make sure to number all your equations so that readers can easily refer to them}

\begin{equation}
\Gamma = \frac{\kappa L}{4\pi c GM}
\end{equation}
where $L$ is the stellar luminosity, has no effect on
the critical rotation because of gravity darkening. In an attempt to clarify the debate, \cite{Maeder1999}
and \cite{MaederMeynet2000} (hereafter referred to as MMM) re-investigated the question and found
two roots to the equation $\bg_{\rm tot}=\boldsymbol{0}$. The first gives the Keplerian angular velocity as
the critical angular velocity for Eddington parameters smaller than $0.639$. The second root yields a critical angular velocity
lower than $\Omega_k$ that tends to zero when the
rotation-dependent Eddington parameter \citep[see][]{Maeder1999} tends to unity for Eddington
parameters larger than $0.639$.

\cite{Maeder1999} based his derivation on the model of \cite{vonzeipel}, which states that the radiative flux $\bF$ at some colatitude $\theta$ on the surface of a rotating star is proportional to the local effective gravity $\bg_{\rm eff}$. For barotropic stars, this leads to
\begin{equation}
\bF= -\rho \chi \frac{dT}{dP}\bg_{\rm eff} \ ,\end{equation}
where $\chi =4 a c T^3 / (3 \kappa \rho)$ is the radiative conductivity. Additionally, assuming solid-body rotation, \cite{zahn92} obtained

\begin{equation}
\rho \chi \frac{dT}{dP}= \frac{L(P)}{4 \pi G M_\star } \ ,
\end{equation}
where 

\begin{equation}
M_\star= M\left( 1 - \frac{\Omega^2}{2 \pi G \rho_{\rm m}}\right) \ ,
\end{equation}
and $\rho_{\rm m}$ is the mean density of the star. $L(P)$ is the luminosity outflowing across the isobar of pressure $P$. \cite{Maeder1999} then wrote the radiative flux in the barotropic case
as

\begin{equation}\label{eq:barotropic}
\bF= - \frac{L(P)}{4 \pi G M_\star } \bg_{\rm eff} \ . 
\end{equation}
In the case of shellular rotation, $\Omega \simeq \Omega(r)$, and
following the work of \cite{zahn92}, \cite{Maeder1999}  linearly developed all quantities
around their average on an isobar and found the radiative flux in
the baroclinic case

\begin{equation}\label{eq:baroclinic}
\bF= - \frac{L(P)}{4 \pi G M_\star } (1-\zeta(\theta))\bg_{\rm eff} \ .
\end{equation}
We note that Eq. (\ref{eq:barotropic}) assumes solid-body rotation, while Eq.
(\ref{eq:baroclinic}) corresponds to the case of slow rotation \cite[][]{zahn92}.
\cite{Maeder1999} noted that $\zeta(\theta)\sim10^{-2}$ so that according
to this model, the ratio $F/g_{\rm eff}$  depends
only mildly on colatitude.

\subsubsection{Interferometric observations}\label{wio}

Recent progresses on rotating stars, both observational and theoretical,  does not confirm this mild dependence, however. Interferometric observations of several rapidly rotating stars \citep[e.g.][]{monnier_etal07,
zhao_etal09, che_etal11, domiciano_etal14} show that if gravity
darkening is modelled by a power law such as

\begin{equation}\label{eq:VZ}
F(\theta) \sim g_{\rm eff}(\theta)^{4 \beta} \ ,
\end{equation}
then $\beta<1/4$ for all the observed stars. Furthermore, the observed exponents
decrease as the rotation rate of the stars increases \cite[e.g.][]{domiciano_etal14}. These results are
in line with the predictions of the ESTER 2D models, which match the observations well
\citep{ELR11}. ESTER 2D models  indeed predict that the relation between flux and
effective gravity is not a power law, but can be approximated as such as a first step. Models also show that $\beta\simeq 0.25$
at low rotation rates, but that $\beta$ decreases to 0.13 when rotation approaches criticality.
This behaviour has implications on the $\Omega\Gamma$-limit introduced
by \cite{MaederMeynet2000}. These limitations prompt us to revisit this limit with ESTER 2D models.

\subsection{The $\omega$-model}

Before using full ESTER 2D models, it is worth considering the problem in light of the simplified $\omega$-model of \cite{ELR11}.
The general purpose of the $\omega$-model is to describe the latitudinal
variations in radiative flux of rotating stars in a simpler way than
with a full 2D model. To this end, it is assumed that the
flux within the radiative envelope of an early-type star can be written  as

\begin{equation}\label{eq:flux_omega}
\bF = -f(r,\theta)\bg_{\rm eff} \ ,
\end{equation}
where $f(r,\theta)$ is some function of the position to be determined.
In this assumption, $\bF$ and $\bg_{\rm eff}$ are assumed to be anti-parallel.
\cite{ELR11} showed that this is a rather good approximation because
full ESTER 2D models show that the angle between the two vectors never
exceeds half a degree, even for the most distorted stars. 

In a radiative stellar envelope,  %\CC{At the stellar surface}
the function $f(r,\theta)$ can be determined
from flux conservation equation, namely \mbox{$\bnabla \cdot {\bF}= 0$}, along
with the assumption that the stellar mass is rather centrally condensed,
so that the Roche model can be used. This implies that the first-order
equation of the flux can be completed by the boundary condition

\begin{equation}
 \lim_{r\rightarrow 0} f(r,\theta) = \frac{L}{4\pi GM} \ .
 \end{equation}

The equation for $f(r,\theta)$ can then be solved analytically
\cite[see][for details]{ELR11,R16a}, with the following result:

\begin{equation}
f(r,\theta)=\frac{L}{4\pi G M}\frac{\tan ^2 \psi(r,\theta)}{\tan^2 \theta} \ ,
\end{equation}
where $\psi(r,\theta)$ is obtained by solving

\begin{equation}\label{eq:psi}
\cos \psi + \ln \tan(\psi/2) = \frac{1}{3}\omega^2r^3 \cos^3 \theta + \cos \theta + \ln \tan(\theta/2) \ .
\end{equation}
In this equation $r$ has been scaled with the equatorial radius $R_{eq}$
and 

\begin{equation}
\omega=\frac{\Omega}{\Omega_k}
\end{equation}
where $\Omega$ is the angular velocity of the star, which is assumed to be uniform (the case of surface differential rotation has been investigated by \citealt{zorec_etal17}). 

At the equator, an analytic expression of $f$ can be obtained,

\begin{equation}
f(r=1,\pi/2)= \frac{L}{4 \pi G M}\left(1-\frac{\Omega^2 R_{eq}^3}{GM}\right)^{-2/3} \ ,
\end{equation}
so that the equatorial radiative flux reads

\begin{equation}\label{eq:eqflux}
\bF(R_{eq},\pi/2)= -\frac{L}{4 \pi  G M}\left(1-\omega^2\right)^{-2/3} \bg_{\rm eff} \ .
\end{equation}

In the slow rotation limit, Eq. (\ref{eq:eqflux}) can be written
\begin{equation}\label{eqfl}
\bF(R_{eq},\pi/2) \simeq  -\frac{L}{4 \pi G M \left( 1 -\frac{2}{3}\omega^2\right)} \bg_{\rm eff} \ .
\end{equation}

In this limit, where $R_{eq} \approx  R_{p}$, this is identical to the MMM expression, which we now obtain for negligible stellar distortions. % \CC{in the $\omega$-model framework.}  

Equations \eqref{eq:eqflux} and \eqref{eqfl} show an important difference: the exact (within the $\omega$-model) expression \eqref{eq:eqflux} shows that the ratio
$F/g_{\rm eff}$ diverges when the
$\Omega$-limit $\omega=1$ is approached, while in its slow rotation approximation \eqref{eqfl}, $F/g_{\rm eff}$ remains finite. This is in line with  von Zeipel's law, which is valid at low rotation rates and which states that flux and effective gravity are proportional. This important difference now calls for a new investigation of the $\Omega\Gamma$-limit.

\subsection{Critical angular velocity at the $\Omega\Gamma$-limit: Ideas from the \om}\label{sec:omgamlim}

\subsubsection{Preliminaries}

With the expression of the radiative flux from the $\omega$-model, we can derive the critical angular velocity
$\Omega_c$ corresponding to the $\Omega\Gamma$-limit. When this limit is reached, then

\begin{equation}\label{eq:gtot0}
\bg_{\rm tot} = \bg_{\rm eff} + \bg_{\rm rad} = \boldsymbol{0} 
\end{equation}
somewhere at the surface of the star. As in MMM, we introduce a limiting flux from Eq. (\ref{eq:grad}) and
the $\Omega\Gamma$-limit condition $\bg_{\rm tot}= \boldsymbol{0}$, namely

\begin{equation}
\bF_{\rm lim}= -\frac{c}{\kappa}\bg_{\rm eff} \ .
\end{equation}
From this expression, we define the rotation-dependent Eddington parameter
$\Gamma_{\Omega}(\theta)$ as the ratio of the actual flux $F(\theta)$
obtained with the $\omega$-model, and the limiting flux, namely

\begin{equation}\label{eq:gamma}
\Gamma_{\Omega}(\theta) = \frac{F(\theta)}{F_{\rm lim}(\theta)} = \frac{\kappa(\theta) }{c} f(r=1,\theta) \ .
\end{equation}
Using Eq. (\ref{eq:gamma}) we can rewrite Eq.
(\ref{eq:gtot}) as

\begin{equation}\label{eq:gtot2}
\bg_{\rm tot} = \bg_{\rm eff} \left[1-\Gamma_{\Omega}(\theta)\right] \ .
\end{equation}
The critical angular velocity $\Omega_c$ is reached if somewhere on
the stellar surface $\bg_{\rm tot}= \boldsymbol{0}$, that is, if there
is a colatitude where either $\Gamma_{\Omega}(\theta) = 1$ or $\bg_{\rm eff}(\theta)=\boldsymbol{0}$.

\subsubsection{Critical latitude: the equator}
In all 2D models both effective temperature and effective gravity are minimum at the equator \citep[e.g.][]{ELR13}. The solution $\bg_{\rm eff}(\theta)=\boldsymbol{0}$ is therefore always reached first at the equator.
We now focus on the $\Gamma_{\Omega}(\theta)=1$ solution.

We first
observe that $\Gamma_{\Omega}(\theta)\propto \kappa(\theta)f(r=1,\theta)$ should be an increasing function of
co-latitude, at least for (very) rapidly rotating stars.  $f(r=1,\theta)$ indeed always increases with $\theta$ and diverges at the equator when the Keplerian angular velocity is approached. As mentioned before, in all 2D models the effective temperature is minimum at equator, but it is not straightforward how to predict whether the opacity $\kappa$ increases or decreases with decreasing $\Teff$ (see appendix \ref{appendixB} for an attempt). Still, we expect the opacity to vary on the stellar surface, but much less than $f(r,\theta)$ near Keplerian angular velocity. According to the \om, and because of the equatorial singularity, it is therefore very likely that the solution $\Gamma_{\Omega}=1$ is always first reached at the equator. %is always maximum at the stellar equator. We thus conclude that the solution $\Gamma_{\Omega}(\theta) = 1$ should also always be reached at equator first. 

\cite{MaederMeynet2000} came to the same conclusion regarding the location of the $\Gamma_{\Omega}(\theta) = 1$ solution on the surface. However, they traced it back to an opacity effect, assuming that the latter increases with decreasing $\Teff$ and thus is highest at the equator. According to ESTER 2D models, this is not the case for rotating stars (see below).

%\corr{Another major discrepancy between \cite{Maeder1999} and \cite{MaederMeynet2000} model for the radiative flux, and the $\omega$-model, is that in the former, the only latitudinal variation of $\Gamma_\Omega$ comes from opacity, if we neglect $\zeta(\theta)$ (which is reasonable according to \citealt{MaederMeynet2000}). Neglecting the latitudinal variations of the opacity would lead \cite{Maeder1999} and \cite{MaederMeynet2000} rotation-dependent Eddington parameter to reach 1 at all latitudes simultaneously, i.e. the complete surface would become unbound. Note that, in this case of constant opacity, not neglecting $\zeta(\theta)$, which is maximum at the poles and decreases towards equator \citep{Maeder1999}, would lead critical rotation to be reached at the poles first.}
%\corr{On the other hand, using the $\omega$-model, the latitudinal dependence of both $f(r,\theta)$ and $\kappa(\theta)$ contribute to the variations of the rotation-dependent Eddington parameter on the stellar surface. Near criticality, $f(r,\theta)$ diverges at the equator and $\Gamma_\Omega$ strongly increases in this region. Therefore, at constant opacity, only a small equatorial region becomes unbound at criticality. When we consider the latitudinal variation of the opacity the latter affirmation remains true because the variations of $f(r,\theta)$ completely dominate those of opacity. According to the $\omega$-model, the opacity variations over the stellar surface are therefore unimportant to determine the latitude where $g_{\rm tot}=0$.}

\subsubsection{Unique critical angular velocity}
\begin{figure}[t]
\centerline{\includegraphics[width=0.5\textwidth]{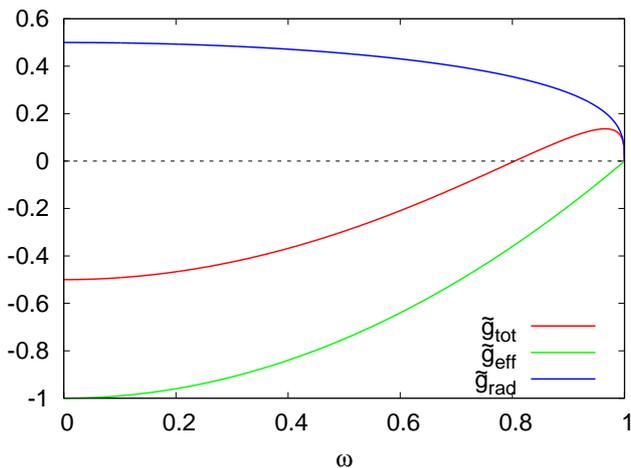}}
  \caption{Scaled total gravity, effective gravity, and radiative acceleration at the equator  as a function of $\omega$ with $\Gamma_{eq}=0.5$. The total gravity at the equator has two zeros; the first root corresponds to $\Gamma_{\Omega}(\pi/2)=1$, and the second to $g_{\rm eff}(\pi/2)=0$.}
  \label{fig:gtot}
\end{figure}

 Equation~(\ref{eq:gtot2}) shows that there are two solutions for 
$\bg_{\rm tot}= \boldsymbol{0}$, and thus two possible critical angular velocities. However, as we show now,
the $\omega$-model removes the $\bg_{\rm eff}= \boldsymbol{0}$ root for the $\Omega\Gamma$-limit and thus points to a single critical angular velocity.  According to the $\omega$-model at equator, Eq. (\ref{eq:gtot}) reads

\begin{equation}
g_{\rm tot}(\pi/2)= g_{\rm eff}(\pi/2)+ g_{\rm rad}(\pi/2) \ ,
\end{equation}
where

\begin{equation}\label{eq:grad_geff}
g_{\rm rad}(\pi/2)=-\frac{\kappa(\pi/2) L}{4 \pi c G M}\left(1-\omega^2\right)^{-2/3}g_{\rm eff}(\pi/2) \ ,
\end{equation}
and

\begin{equation}
g_{\rm eff}(\pi/2) = - R_{eq}\Omega_k^2 \left(1 -
\omega^2 \right) \ .
\end{equation}
Here $\geff$ and $g_{\rm rad}$ are the radial components of the accelerations (thus positive when outwards). We can then  write the equatorial
total gravity scaled with $\Omega_k^2R_{eq}$ as

\begin{equation}
\tilde{g}_{\rm tot}(\pi/2) = \omega^2+\Gamma_{eq} (1-\omega^2)^{1/3}-1 \ ,
\end{equation}
where $\Gamma_{eq}$ is the standard Eddington parameter evaluated at the equator. From Eq. (\ref{eq:grad_geff}), we
see that at the equator, the ratio $g_{\rm rad}/g_{\rm eff}$   increases as $(1-\omega^2)^{-2/3}$ with increasing $\omega$, which also implies that if $g_{\rm eff}$
approaches $0$ when $\omega\tv1$, $g_{\rm rad}$
will also tend to $0$ but more slowly. Figure \ref{fig:gtot} shows the
scaled total gravity, effective gravity, and radiative acceleration at
the equator as a function of $\omega$ with $\Gamma_{eq}=0.5$. The total
gravity at the equator has two zeros; the first root corresponds to
$\Gamma_{\Omega}(\pi/2)=1,$ and
the second root gives $g_{\rm eff}(\pi/2)=0$.

For sub-critical rotation (i.e. $g_{\rm tot} < 0$  or $|g_{\rm eff}|>g_{\rm rad}$), the star is gravitationally bound. When we increase $\omega$, the equatorial effective gravity $|g_{\rm eff}|$
decreases faster than $g_{\rm rad}$, to the point where $|g_{\rm
eff}|=g_{\rm rad}$ (equivalently, $\Gamma_{\Omega}(\pi/2)=1$), at this
point, $\Omega=\Omega_c$ and $g_{\rm tot}=0$. Increasing $\omega$ even more would result in a radiative acceleration that
surpasses the effective gravity at equator. When this happens, $g_{\rm
tot}>0$ and the star is no longer gravitationally bound up to $\omega=1$
where the second root is reached. The solution $\Gamma_{\Omega}(\pi/2)=1$
is therefore \textit{\textup{always}} reached before $\omega=1$, when evolution (say) drives the growth of $\omega$.

\subsubsection{Critical rotation given by the \om}

The main difference between the MMM model and ours, in addition to our unique critical angular velocity, comes from the latitudinal variation of $\Gamma_\Omega$. In MMM models the latitudinal variations in $\Gamma_\Omega$ come from the latitudinal variations in opacity when we discard the small correcting function $\zeta(\theta)$. As a consequence, if the surface opacity were constant (e.g. with Thomson opacity of electrons), $\Gamma_\Omega$ would reach unity at all latitudes at the same time when $\omega$ increases! The \om\ predicts that the ratio between effective gravity and radiative flux depends on latitude and diverges at the equator when criticality approaches. Even in the extreme case of a constant surface opacity, only a small equatorial region therefore becomes unbound at criticality. The \om\ shows that opacity variations over the stellar surface are unimportant for determining the latitude where $g_{\rm tot}=0$ because of the equatorial singularity. This discussion demonstrates that the use of a constant ratio between the surface flux and the effective gravity (the von Zeipel law) as done in the MMM model has an important consequence for determining a critical rotation because it removes the equatorial singularity of the ratio $\Teff^4/\geff$.

In line with the $\omega$-model and the maximum of $\Gamma_\Omega$ at equator, the condition giving the critical angular velocity $\Omega_c$ is

\begin{equation}
\Gamma_\Omega(\pi/2)=\frac{\kappa(\pi/2)L}{4
\pi c G M}\left( 1- \frac{\Omega_c^2}{\Omega_k^2}\right)^{-2/3}
= 1 \ ,
\end{equation}
or equivalently,

\begin{equation}\label{eq:omegac}
\Omega_c=\Omega_k \sqrt{1-\Gamma_{eq}^{3/2}}
\ .
\end{equation}
%where

%\begin{equation}
%\Gamma_{eq} = \frac{\kappa(\pi/2)L}{4 \pi c G M}
%\end{equation}
%is the standard Eddington parameter evaluated at equator.

These equations show that $\Omega_c$ is reduced with increasing Eddington parameter compared to $\Omega_k$. \cite{MaederMeynet2000} came to the same conclusion, but with a different expression for critical angular velocity, namely $\Omega_c \propto \Omega_k \sqrt{1-\Gamma_{eq}}$. Because $\Gamma_{eq} \leq 1$, their ratio critical to Keplerian angular velocity is lower for the same equatorial Eddington parameter.  When radiative acceleration effects are weak at the equator, that is, when $\Gamma_{eq} \ll 1$, we find $\Omega_c \simeq \Omega_k$, as expected. This is also the solution of \cite{MaederMeynet2000} for critical angular velocity in this regime. 
%Similarly, both this work and \cite{MaederMeynet2000} find $\Omega_c$  to vanish if $\Gamma_{eq}$ tends towards unity.

\subsubsection{Some conclusions from the \om}

The analysis based on the \om\ underlines three important points:

\begin{enumerate}
\item Formally, the Keplerian angular velocity is never reached. The critical angular velocity such that the centrifugal acceleration overcomes the sum of the gravitational and radiative accelerations at some place on the stellar surface is always lower than the Keplerian angular velocity.
\item This balance of forces is always first reached at the equator when $\Omega/\Omega_k$ increases because the ratio $\Teff^4/\geff$ at the equator diverges when criticality is reached.
\item The use of the von Zeipel law, which assumes the proportionality of the surface flux with the effective gravity gives a critical latitude that depends on the latitudinal variations of the surface opacity and is therefore not necessarily located at the equator.
\end{enumerate}
These conclusions based on the \om\ immediately raise the question of the accuracy of this model. This is the next point that we discuss in light of observations and full 2D ESTER models.

%\corrM{not here mais à caser qqpart}: "However, the foregoing results are within the $\omega$-model assumptions, namely that, regardless of the rotation rate, \sout{the radiative flux vector is anti-parallel to the effective gravity and that} massive and intermediately massive stars have an internal mass distribution that can be represented by a Roche model. These assumptions are never exactly satisfied, especially if we tackle the critical angular velocity. \corr{Still, a comparison between ESTER 2D-calculations and results using the $\omega$-model shows good agreement for stars with rotation rates up to 90~$\%$ of the Keplerian angular velocity, see appendix~\ref{appendixC}.} "

\subsection{The $\Omega\Gamma$-limit with ESTER 2D models}

\subsubsection{Interferometric observations (again)}

We first briefly return to observations. As described above (\S\ref{wio}), interferometric observations of fast-rotating stars all show that $\beta<1/4$ when the surface flux distribution is assumed to vary as $\Teff\propto\geff^\beta$. Moreover, they clearly show that $\beta$ decreases with increasing rotation \citep{domiciano_etal14}. From this discussion we can now interpret the fact that $4\beta<1$ and decreases with increasing rotation as evidence for a divergence of the ratio $\Teff^4/\geff$ at the equator when criticality is approached.

\subsubsection{Accuracy of the $\omega$-model}

The assumptions of the \om\ are that the flux vector is anti-parallel to the effective gravity, the gravitational field is that of a point mass (the Roche model), and the rotation is uniform. The last two of these approximations probably entail the largest errors. They can be appreciated by comparing the flux latitudinal distribution of the \om\ with the output of 2D ESTER models. A comparison has been made in \cite{ELR11}, but here we focus on the relative difference between the flux of the two models.

\begin{figure}[t]
\includegraphics[width=0.5\textwidth]{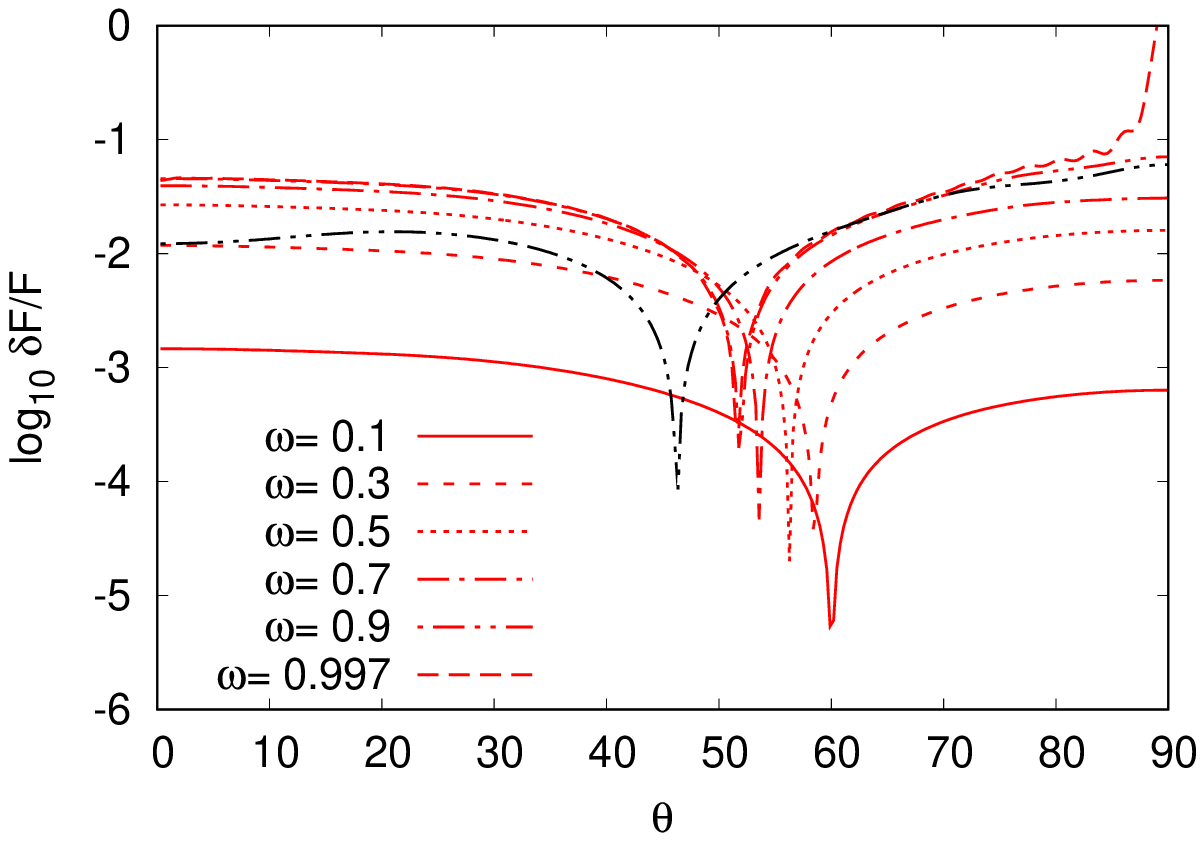}\hfill
\includegraphics[width=0.5\textwidth]{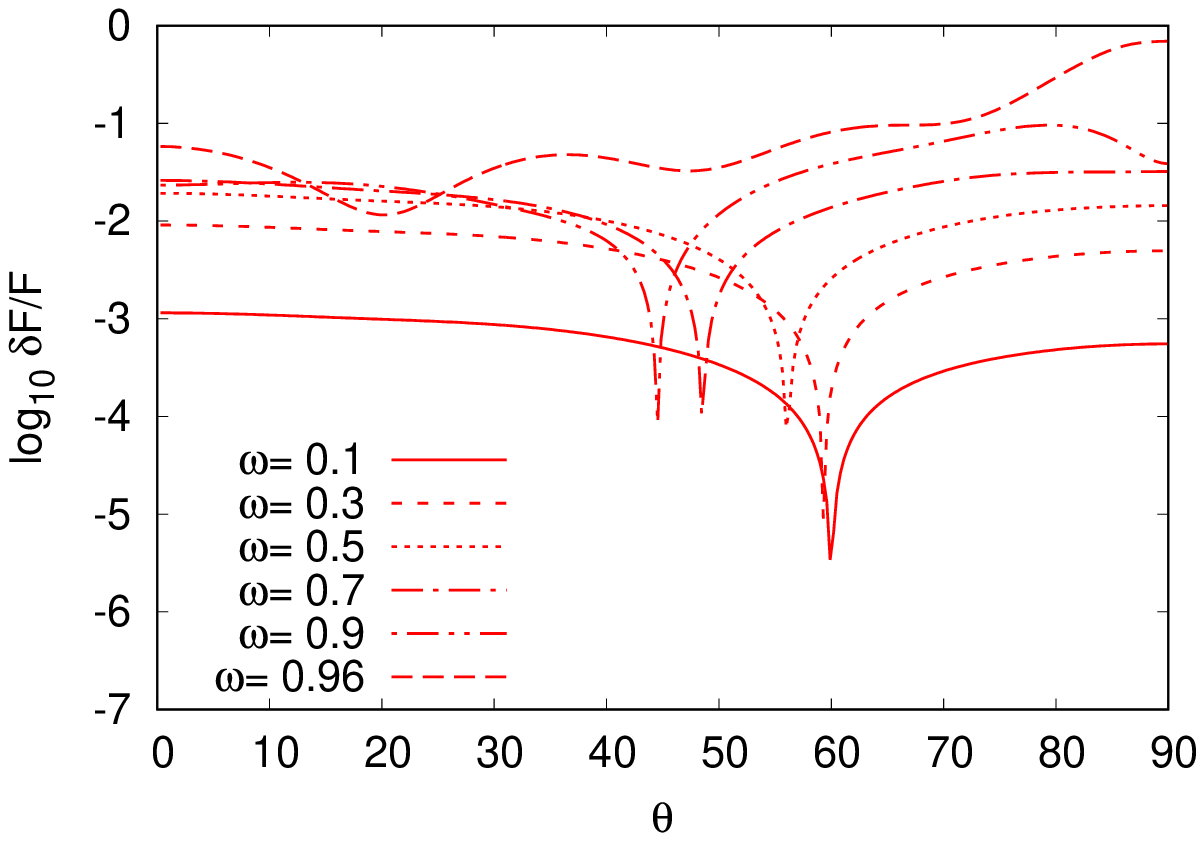}
  \caption{Relative difference between the radiative flux of the $\omega$-model and the ESTER model as a function of co-latitude for a $15~M_\odot$ (top) and a $40~M_\odot$ (bottom) ZAMS-star with various angular velocity ratios. The black line corresponds to an evolved $15~M_\odot$ ESTER 2D model with $\omega=0.9$ and a fractional abundance of hydrogen in the convective core $X_{\rm core}/X_0=0.5$. $X_0$ is the initial hydrogen mass fraction at ZAMS. The minimum of each curve corresponds to a sign change of $F_{\rm ESTER} -F_\omega$.}
  \label{fig:A1}
\end{figure}

We computed the flux from the $\omega$-model, Eq.~(\ref{eq:flux_omega}), where $L$, $M$, $\omega$, $r,$ and $g_{\rm eff}$ were taken from the output of ESTER 2D models. For two 2D ESTER ZAMS models of 15~\msun\ and 40~\msun, we computed the relative difference between the fluxes of the ESTER and \om, namely

\begin{equation}
\frac{\delta F}{F}= \frac{|F_{\rm ESTER} -F_\omega|}{F_{\rm ESTER}}
\end{equation}
as a function of co-latitude. The result is shown in Fig.~\ref{fig:A1}.

On the ZAMS, and for rotation rates of up to 90~\% of the Keplerian angular velocity, the relative difference between the fluxes remains lower than $10~\%$. For angular velocity ratios lower than $50~\%,$ this difference drops to less than one percent, making the $\omega$-model quite reliable for most of the rapidly rotating stars.

This comparison has been made at ZAMS. As stars evolve along the MS, they become more and more centrally condensed (this is discussed in the follow-up paper) and thus better satisfy the Roche approximation. Therefore, the relative deviation between the radiative flux of ESTER 2D models and the analytic $\omega$-model probably never exceeds 10~\% for stars with $\omega \leq 0.9$ during the MS. This is illustrated in Fig.~\ref{fig:A1} (top) with a 15~\msun\ model at mid-MS rotating with $\omega=0.9$. Clearly, the relative difference is reduced compared to the ZAMS model.

\subsubsection{The $\Omega\Gamma$-limit with ESTER 2D models}

The current ESTER 2D models describe the steady state of a rotating star with a convective core and a radiative envelope, that is, an early-type star. Compared to previous attempts of making stellar models in two dimensions \citep[e.g.][]{roxburgh04, JMS05}, ESTER models self-consistently include the differential rotation of the radiative envelope that is driven by the baroclinic torque. They also treat self-consistently the associated meridional circulation. A brief description of these models is given in Appendix B, but we refer to the original papers of \cite{ELR13} and \cite{Rieutordal2016} for a more detailed account.

\begin{figure}[t]
\includegraphics[width=0.5\textwidth]{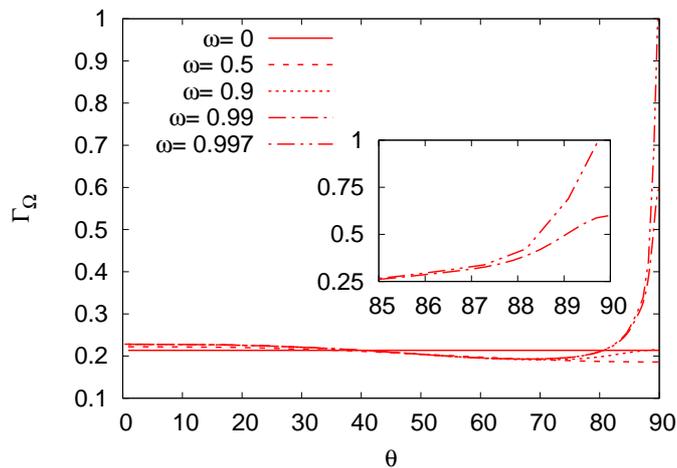}
\includegraphics[width=0.5\textwidth]{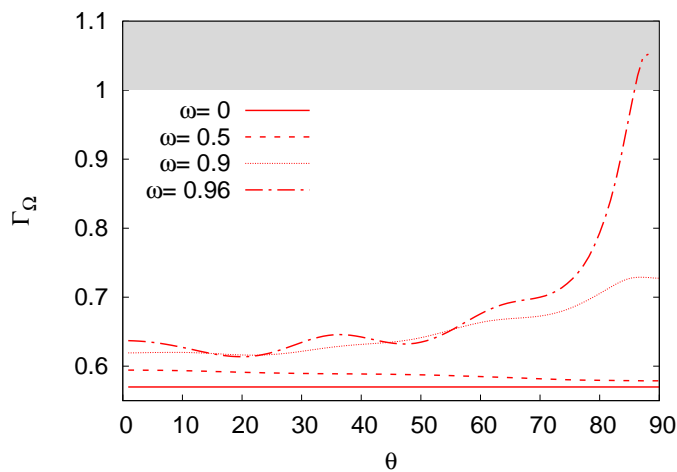} 
\caption{Rotation-dependent Eddington parameter $\Gamma_\Omega(\theta)$ as a function of colatitude for various fractions of the Keplerian angular velocity for a $15~M_\odot$ (top) and a $40~M_\odot$ (bottom) ESTER model at ZAMS, with $Z=0.02$. The grey area corresponds to supercritical rotation.}
  \label{fig:gamma}
\end{figure}

In Figure~\ref{fig:gamma} we illustrate the latitudinal variations
in $\Gamma_{\Omega}$ for a
$15~M_\odot$ ESTER model and for a $40~M_\odot$ ESTER model both taken at ZAMS, computed for the metallicity $Z=0.02$ and for various values of $\omega$ that we now define as $\omega= \Omega_{eq}/\Omega_k$ (see below). 

We first consider the $15~M_\odot$ ESTER-model at ZAMS. Interestingly, we find that $\Gamma_{\Omega}(\theta)$ first slightly decreases with increasing co-latitude before eventually vigorously increasing near equator at high angular velocity ratios. The decrease in $\Gamma_{\Omega}(\theta)$ is clearly an opacity effect, which we trace back to the density decrease with $\theta$ along the stellar surface. The increase near equator at high rotation speeds is an effect of the divergence of the function $f(r=1,\theta)$. Surprisingly, we see that $\Gamma_{\Omega}=1$ requires $\omega=0.997,$ showing that the difference between the actual critical angular velocity and the Keplerian velocity is really tiny for a $15~M_\odot$ ZAMS star. 

To strengthen the effects of radiative acceleration, we considered the case of a $40~M_\odot$ ZAMS model. Here we also see (Fig.~\ref{fig:gamma} bottom) that $\omega$ must be as high as $\sim 0.96$ for the $\Omega\Gamma$ limit to be reached\footnote{ $\omega$ has to be slightly lower than $0.96$ so that $\Gamma_\Omega$ is exactly unity at equator. At $\omega=0.96,$ the star is already supercritical at the equator (grey area in Fig.~\ref{fig:gamma}).}. Technically, these latter results are not as precise as those for the 15~\msun\ model because we approach the current limits of the ESTER code in terms of resolution, but they also point to a small difference between $\Omega_c$ and $\Omega_k$.

%\corr{According to the $\omega$-model
%\begin{equation}
%\Gamma_\Omega(\theta)=\Gamma(\theta) \frac{\tan^2 \psi(r=1,\theta)}{\tan^2 \theta}
%\end{equation}
%}

\begin{figure}[t]
    \includegraphics[width=0.5\textwidth]{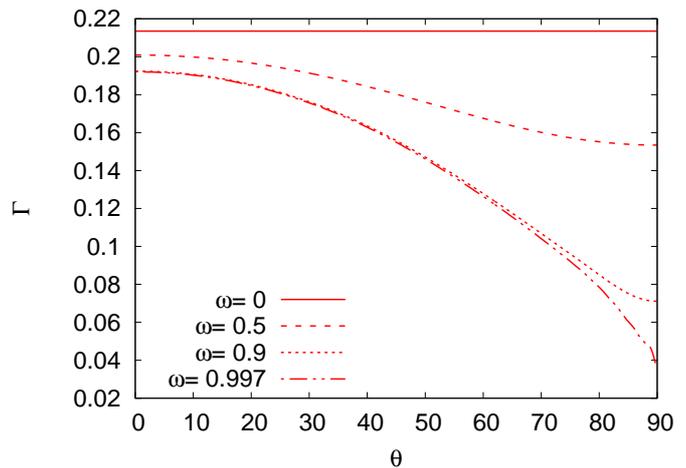}\hfill
    \includegraphics[width=0.5\textwidth]{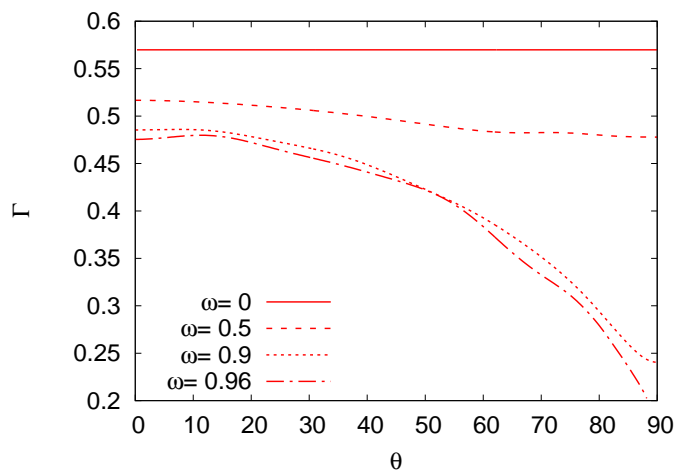}
    \caption{Eddington parameter $\Gamma$ as a function of co-latitude for a 15~$M_\odot$ (top) and a 40~$M_\odot$ ESTER 2D model (bottom) at ZAMS and for various angular velocity ratios, with $Z=0.02$.}
    \label{fig:Gamma_theta}
\end{figure}

From Eq.~(\ref{eq:omegac}), at ZAMS, we find the equatorial Eddington parameter $\Gamma_{eq} \simeq 0.033$ at criticality for the $15~M_\odot$ ESTER model and $\Gamma_{eq} \simeq 0.18$ for the  $40~M_\odot$ ESTER model. This is surprisingly low for such massive stars. To clarify this result, Fig.~\ref{fig:Gamma_theta} shows the latitudinal variations in Eddington parameter $\Gamma$ for both the $15~M_\odot$ and $40~M_\odot$ ESTER 2D models and for various angular velocity ratios. For the two models, $\Gamma$ decreases with co-latitude when rotation is non-zero. The more rapid the rotation, the lower $\Gamma_{eq}$. The only latitudinal dependence of the Eddington parameter being on opacity, we trace back the decrease in the latter at low latitudes to the decrease in surface density (see Appendix~\ref{appendixB}). We note that the opacity cannot be lower than a minimum set by pure electron scattering. While the latitudinal variations of $\kappa$ are somewhat unimportant for determining the spatial location of criticality, they are crucial to the value of $\Omega_c/\Omega_k$. 

\begin{figure}[t]
    \includegraphics[width=0.5\textwidth]{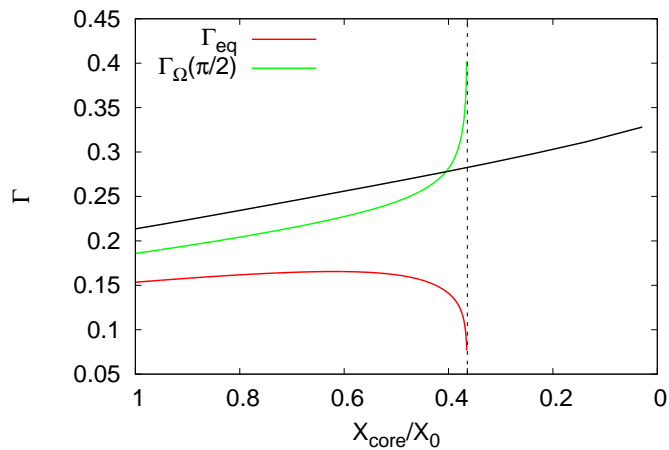}
    \caption{$\Gamma_{eq}$ and $\Gamma_{\Omega}(\pi/2)$ as a function of the fractional abundance of hydrogen in the convective core for a $15~M_\odot$ ESTER 2D model initially rotating at $\omega_i=0.5$. The  black line corresponds to the evolution of $\Gamma_{eq}$ for $\omega=0$.}
    \label{fig:Gamma_evol}
\end{figure}

It might be wondered, however, whether more evolved or more massive stars might have larger $\Gamma_{eq}$ and thus a critical angular velocity that is farther from the Keplerian angular velocity. Fig.~\ref{fig:Gamma_evol} shows the evolution of both $\Gamma_\Omega(\pi/2)$ and $\Gamma_{eq}$ as a function of the fractional abundance of hydrogen in the convective core $X_{\rm core}/X_0$ for a $15~M_\odot$ ESTER 2D model initially rotating at $\omega_i \equiv \Omega_{eq,i}/\Omega_k = 0.5$, and without considering any mass loss. For the non-rotating case, evolution tends to  increase $\Gamma_{eq}$ . The increase in luminosity associated with nuclear evolution surpasses the decrease in surface opacity that is due to stellar expansion. However, when rotation is included and $\omega_i=0.5$, criticality is reached when $X_{\rm core}/X_0 \simeq 0.36$. At this time, $\Gamma_\Omega$ diverges. While evolution proceeds and $\omega$ grows, the star flattens considerably, causing a significant drop in opacity in the equatorial region. This is clearly shown by the $\Gamma_{eq}$ curve of Fig.~\ref{fig:Gamma_evol}. After a slight increase at the beginning of evolution, $\Gamma_{eq}$ drops when criticality approaches. For this model, the equatorial opacity reduction completely dominates the effect of luminosity growth due to evolution.

%Our interpretation is that, as $\omega$ increases towards criticality, the star flattens and both the equatorial effective temperature and density decrease, thus making the opacity decrease. It seems that this opacity reduction completely dominates the increase of luminosity caused by nuclear evolution, and thus solely dictates the evolution of $\Gamma_{\rm max}$. We are not able to reach $\Gamma_{\Omega}=1$ because of the singularity, but we see that $\Gamma_{\rm max}$ has already decreased to a value close to what we obtained for criticality at ZAMS (i.e. $\sim 0.04$).}

%\corr{
%In conclusion, we find the interpretation if this ``mild'' difference between critical and Keplerian angular velocities to be rather difficult due to the non-linearity of the processes involved and the singularity of the critical rotation. In order to better understand the strong decrease of $\Gamma_{\rm max}$ leading to the mild difference between the $\Omega_c$ and $\Omega_k$, it would be useful to analyse the equatorial singularity near critical rotation with a dedicated 2D atmospheric model. This is however beyond the scope of the present work. }

These results tend to confirm the idea put forward by \cite{Glatzel1998}, namely that the critical angular velocity is not strongly modified  by the Eddington limit. We may conclude that because of the noticeable effect of rotation on opacity at the equator, the critical angular velocity is only slightly reduced compared to the Keplerian angular velocity, at least for stars with a mass lower than $40~\msun$ at $Z=0.02$.

In practice, the difference is therefore tiny enough to be neglected in view of the other uncertainties of stellar models. We therefore continue to express $\Omega$ as a fraction of the equatorial Keplerian angular velocity $\Omega_k$ to appreciate the distance to criticality, bearing in mind that this fraction is slightly smaller than the actual one.

\section{Local mass-flux prescription from 1D models}\label{sec:search}

We now address the second question of the paper: the dependence of mass and angular momentum losses on rotation rate.
All hot stars have radiation-driven winds that become directly
observable in spectral energy distributions and spectral lines as soon as
they are above some luminosity threshold. For massive stars of spectral
types O, B, and A, this threshold corresponds to $L \sim 10^4L_\odot$ \citep{Abbott1979}. Above this luminosity, massive stars show direct spectroscopic evidence of winds throughout their lifetime (UV P Cygni line profiles and optical emission lines such as H$\alpha$, see \citealt{Abbott1979} or \citealt{Kudritzki2000} and references therein). The radiative acceleration has a significant effect on the total gravity, and as shown in the previous section, reduces the critical angular velocity.

In this section we propose to derive a local mass-flux prescription that can be seen as a local equivalent of CAK original theory where the radiation-driven wind is assumed to be an isothermal stationary flow that is driven outward by photon scattering and absorption. We account for the finite cone angle of the radiating photospheric surface and for radial variations of ionisation using the results of \cite{PPK} and \cite{Friend1986}.
Still, with or without these additional corrections, the global mass-loss rate  follows a similar scaling for the wind momentum-luminosity relation \citep[][]{kudritzki1995, Puls96}. 

\subsection{Global mass-loss rate derived from 1D CAK theory}

%\subsubsection{Mass-loss rate prescription from CAK}

In the 1D spherically symmetric case (i.e. without rotation), the two hydrodynamical equations needed to describe the mass-flux are the conservation of mass,

\begin{equation}
\dot{M} = 4 \pi r^2 \rho v = {\rm Cst} \ ,
\end{equation}
and the radial momentum equation,

\begin{equation}
v \frac{\partial v}{\partial r} = -\frac{1}{\rho} \frac{\partial p}{\partial r} + g + g_{\rm rad} \ ,
\end{equation}
where $\dot{M}$ is the total mass-loss rate of the star, and $\rho$, $v$, $p$ are the density, radial velocity, and gas pressure, respectively. $g_{\rm rad}$ is the radiative acceleration,

\begin{equation}
g_{\rm rad} = g_{\rm rad}^{\rm line} + g_e \ ,
\end{equation}
where $g_{\rm rad}^{\rm line}$ is the line-driven acceleration, and $g_e= \kappa_e F / c$ is the radiative acceleration due to Thomson scattering. $\kappa_e$ is the opacity from electron scattering. Because we are interested in O, B, and A stars, both bound-free and free-free transitions are neglected \citep[e.g.][]{Runacres1994, Gayley1995}. This may not be valid for Wolf-Rayet stars, however. Finally, $g$ is the gravitational acceleration,

\begin{equation}\label{geff}
g = -\frac{GM}{r^2} \ .
\end{equation}

Using the ideal gas equation of state, we write

\begin{equation}\label{EOS}
p=c_s^2 \rho \ ,
\end{equation}
where $c_s$ is the isothermal sound speed.

Because winds from hot stars are mostly line driven, the evaluation of the line-driven radiative acceleration plays a crucial role in determining the mass flux.  In the Sobolev approximation (i.e. large velocity gradient approximation), considering a purely radial streaming radiation from a point-source star, the line-driven radiative acceleration can be written (CAK)

\begin{equation}
g_{\text{rad}}^{\rm line}= M(t) g_e \equiv k  \left(\frac{\partial v/\partial r}{\rho v_{\rm th} \kappa_e}\right)^\alpha\ g_e \ ,
\end{equation}
where $M(t)= k t ^{-\alpha}$ is the CAK force multiplier. $\alpha$ and $k$ are the CAK force multiplier parameters (FMPs). $\alpha$ can be interpreted as the ratio of the line force from optically thick lines to the total line-force, which thus decreases with decreasing effective temperature because of the increased iron group lines \cite[e.g.][]{Puls2000}. Moreover, the quantity $k$  is related to the fraction of the total stellar flux, which would be blocked in the photosphere if all lines were optically thick \citep{Puls2000}. $t$ is the electron optical depth parameter and $v_{\rm th}$ is the thermal speed, usually taken as the proton thermal speed  $v_{\rm th}\equiv (2k_{\rm B}T_{\rm eff}/m_{\rm H})^{1/2}$ \citep[e.g.][]{Abbott1982}. However, at least in the lower part of the wind, Fe line-driving dominates. Therefore we instead take $v_{\rm th}\equiv (2k_{\rm B}T_{\rm eff}/m_{\rm Fe})^{1/2}$ for the standard CAK formalism.
We then obtain the global mass-loss rate of a non-rotating star, namely

\begin{equation}\label{eq:CAK}
\dot{M}_{\rm CAK} = \frac{4 \pi}{\kappa_e v_{\rm th}} \left(\frac{k \alpha \kappa_e L}{4 \pi c}\right)^{1/\alpha} \left( \frac{1-\alpha}{\alpha}\right)^{\frac{1-\alpha}{\alpha}}\ [GM(1-\Gamma_e)]^{ \frac{\alpha-1}{\alpha}}.
\end{equation}
The global mass loss (without rotation) thus scales as
\begin{equation}
\dot{M}_{\rm CAK} \propto \left[M(1-\Gamma_e)\right]^{(\alpha-1)/\alpha}L^{1/\alpha} %\ , 
,\end{equation}
which is the basis of the wind momentum-luminosity relation \citep[][]{kudritzki1995, Puls96}.

%\cite{Gayley1995} introduced an alternative formalism that, within the scope of our work, is equivalent to the standard CAK formalism (see \citealt{Puls2000}). %We give a short account of this approach in appendix C.

\subsection{Finite disc and ionisation corrections}

In the CAK approach, the purely radial streaming radiation leads to an electron optical depth parameter $t$ that only depends on $(dv/dr)^{-1}$. This assumption neglects the finite cone angle of the radiating photospheric surface, however. Using Eq.~(49) of CAK, we may rewrite $t$ with its exact expression, leading to the modified force multiplier \citep{PPK}, namely

\begin{equation}\label{eq:modifgrad}
M(t') = M(t) \frac{2}{1-\mu_\ast} \int_{\mu_\ast}^1 \left[ \frac{(1-\mu^2) v/r+\mu^2 v'}{v'}\right]^\alpha \mu d\mu \ ,
\end{equation}
where $v'=dv/dr$, $u = -R/r$, and  $\mu$ is the cosine of the angle between the direction of emitted radiation  and the radial direction and $\mu_\ast = \sqrt{1 - u^2}$. Evaluating the integral in Eq. (\ref{eq:modifgrad}) yields the modified force multiplier, corrected for finite cone angle, namely

\begin{equation}\label{eq:modifgrad2}
\begin{aligned}
M(t') \simeq  \frac{M(t)}{u^2(1+\alpha)(1+\frac{w}{uw'})}\left[ 1 - \left(1-u^2-u\frac{w}{w'} \right)^{1+\alpha}\right] \ ,
\end{aligned}
\end{equation}
where $w=v/v_{\rm th}$ and $w'= dw/du$. As in the original CAK derivation, the mass-loss rate is calculated at the critical radius $r_c$ defined by a singularity and a regularity condition. Following  \cite{PPK}, we assumed $r_c$ to be located very close to the stellar radius, $r_c \simeq R$. This assumption may not be verified for rotators close to criticality, for which the fast-wind solution (with $r_c \simeq R$) is replaced by the so-called $\Omega$--slow solution in the equatorial plane. This solution is characterised by an increased mass-loss rate, a slower and denser wind with a critical radius that is much farther in the wind \citep{Cure2004, Cure2005, Araya2017}. Because the increase in mass loss associated with the $\Omega$--slow solution was quite modest (factor $\sim 2$ in \citealt{Cure2004}), we decided to ignore it.  We further assumed the velocity to follow a power law like

\begin{equation}
v(u)=v_\infty(1+u)^\beta \ ,
\end{equation}
where $v_\infty$ is the terminal velocity of the wind and $0.7\infapp\beta \infapp 1.3$. The corrected force multiplier then simplifies to

\begin{equation}\label{eq:modifgrad3}
M(t') \simeq \frac{M(t)}{1+\alpha} \ ,
\end{equation}
and results in a modified prefactor for the mass-loss rate. Thus we use

\begin{equation}\label{eq:dotM}
\dot{M}= \left(\frac{1}{1+\alpha}\right)^{1/\alpha} \dot{M}_{\rm CAK} \ .
\end{equation}

Because $0<\alpha<1$, the finite-disc correction reduces the global mass-loss rate, and for typical values of $\alpha$, namely between 0.4 and 0.7, the CAK mass-loss rate is multiplied by a factor $\sim 4/9$. 
Additionally, the effect of radial changes in ionisation in the outward direction in the wind can be approximately taken into account by correcting the force multiplier of Eq. (\ref{eq:modifgrad3}), namely multiplying it by a factor $(n_e/W)^\delta$ \citep{Abbott1982}, where $n_e$ is the electron density in units of $10^{11} \rm cm^{-3}$ and $W \equiv 0.5 (1-\sqrt{1-u^2})$ is the radiation dilution factor. $\delta$ is then another FMP. This modification of the line-driven acceleration can be roughly accounted for by replacing  $\alpha$ in the power exponents of Eqs. (\ref{eq:CAK}) and (\ref{eq:dotM}) with $\alpha' \equiv \alpha - \delta$ \citep{Puls96, Puls2000}. Finally, we obtain the modified local mass-flux in the non-rotating case, 

\begin{equation}\label{eq:1}
\begin{aligned}
\dot{m} \equiv  \frac{\dot{M}}{4 \pi R^2}=&\left(\frac{\alpha}{ v_{\rm th} c}\right)\left(\frac{k}{1+\alpha}\right)^{1/\alpha'}  \\
 \times &  \left[ \frac{c}{\kappa_e (1-\alpha)}\left(|g|- \frac{\kappa_e F}{c}\right)\right]^{\frac{\alpha'-1}{\alpha'}}  F^{1/\alpha'}\ ,
\end{aligned}
\end{equation}
where we used the radiative flux $F$ rather than the luminosity. %Alternatively, with the $\overline{Q}$-formalism we obtain

%\begin{equation}\label{eq:2}
%\begin{aligned}
%\dot{m}_2 \equiv \frac{\dot{M}_2}{4 \pi R^2}=& \left(\frac{1}{1+\alpha}\right)^{1/\alpha'}\frac{\alpha}{c^2(1-\alpha)}  \\
% \times &  \left[ \frac{c}{\kappa_e \overline{Q}}\left(g- \frac{\kappa_e F}{c}\right)\right]^{\frac{\alpha'-1}{\alpha'}} F^{1/\alpha'}\ .
%\end{aligned}
%\end{equation}

%Including deviations from the point-like star approximation would result in a modified pre-factor, and the inclusion of ionisation effects would be obtained modifying the $\alpha$ parameter in the exponents by $\alpha'= \alpha - \delta$ where $\delta$ is the CAK ionisation parameter of order $0.1$ for typical O-stars.

 Unlike the approach of MMM, we do not need to express the mass flux so that it explicitly depends on the total gravity. Rather, it now depends on gravity $g$, corrected for the radiative acceleration \textit{\textup{from electron scattering}} $\kappa_e F/c$. %Doing so, we avoid the confusion implying that the relevant opacity should be $\kappa_e$ instead of the total opacity $\kappa$. 
 
 \subsection{Parametrisation of the FMPs}\label{sec:param}

We now focus on the different FMPs $\alpha$, $k,$ and $\delta$ to estimate how they vary with the effective temperature $T_{\rm eff}$.
We assumed that $\delta$ does not significantly vary with $T_{\rm eff}$ and took $\delta = 0.1$, a typical value for hot stars at solar metallicity \citep{Abbott1982}. We note that  $\delta$ can reach much higher values in very metal-poor stars where the wind is mostly driven by hydrogenic lines, and can even be negative under very specific conditions  \citep[][]{Puls2000}. 
 
\begin{figure}[t!]
\centerline{\includegraphics[width=0.5\textwidth]{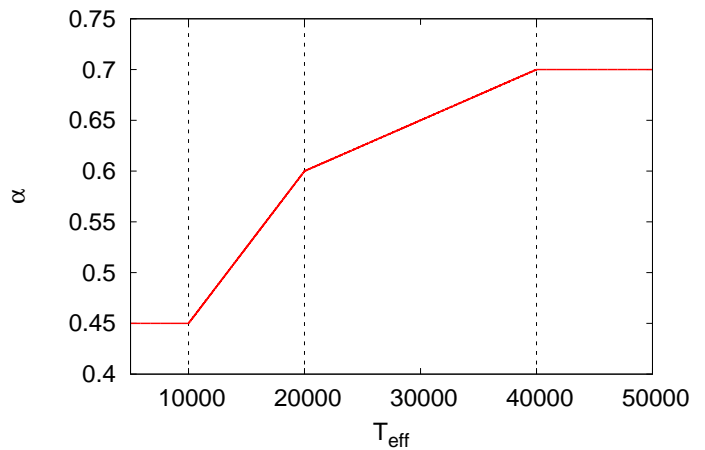}}
  \caption{Adopted force multiplier parameter $\alpha$ as a function of the effective temperature $T_{\rm eff}$ from Eq. (\ref{alpha}) (courtesy J. Puls). The black vertical dashed lines mark the location of the imposed values for $\alpha$.}
  \label{fig:alpha}
\end{figure}

For $\alpha,$  we took fixed values at $T_{\rm eff}= 10~{\rm kK}, \ 20~{\rm kK}, \ 30~{\rm kK,} \ {\rm and} \ 40~{\rm kK}$ and imposed linear interpolation in between (J. Puls priv. comm.), namely

\begin{equation}\label{alpha}
\alpha(T_{\text{eff}}) =
\begin{cases}
    0.45, & \text{if $T_{\text{eff}} \leq 10~{\rm kK}$ \ ,}\\
    1.5\times 10^{-5} T_{\rm eff} + 0.3, & \text{if $10\rm{kK} <  T_{\text{eff}} \leq 20~{\rm kK}$ \ ,}\\
        5\times 10^{-6} T_{\rm eff} + 0.5, & \text{if $20\rm{kK} <  T_{\text{eff}} \leq 40~{\rm kK}$ \ ,}\\
            0.7, & \text{if $T_{\text{eff}} > 40~{\rm kK}$ \ .}\\
  \end{cases}
\end{equation}
This function is shown in Fig.~\ref{fig:alpha}.

 Finally, we calibrated $k$ assuming that our expression for the mass-loss rate  in the non-rotating regime  of Eq. (\ref{eq:1}) is equivalent to the expression of \cite{Vink2001}. The \citet{Vink1999} calculations of wind models for OB stars
showed that around $T_{\text{eff}} \simeq 25~{\rm kK}$, the mass-loss rate
$\dot{M}$ suddenly increases (towards lower
$\Teff$) as a result of the recombination of Fe IV into Fe III, which has a stronger line acceleration in the lower part of the wind.  \cite{Lamers1995} and  \citet{Vink1999} suggested the existence of a second bi-stability jump, around $T_{\text{eff}} = 10~{\rm kK}$, that would be caused by the recombination of Fe III into Fe II. \cite{Vink2001} did not account for this jump, however.

 The \cite{Vink2001} prescription for mass loss still awaits confirmation, however. Their predictions for the size and position of the main bi-stability jump have not been confirmed by observations until today. For instance, \cite{Markova2008} found an $\dot{M}$ jump of a factor in between $0.4$ and $2.5$, and more recent theoretical modelling by \cite{Petrov2016} found the bi-stability jump at $T_{\rm eff}^{\rm jump} \simeq 20~{\rm kK}$, while \cite{Vink2001} predicted an $\dot{M}$-jump by a factor $\sim 10$ located at ${T_{\rm eff}^{\rm jump}} \simeq 25~{\rm kK}$. \cite{Crowther2006}, on the other hand, found a more gradual decrease in terminal velocity $v_\infty$ instead (thus a more gradual increase in $\dot{M}$). In addition, a discrepancy of a factor $2$--$3$ also appears when the mass-loss rates of hot OB stars are compared with $T_{\rm eff} > T_{\rm eff}^{\rm jump}$ obtained with the \cite{Vink2001}  models and from X-ray, UV, and IR diagnostics \citep[e.g.][]{Najarro2011,Sundqvist2011,Bouret2012,Cohen2013, Leutenegger2013, Herve2013, Rauw2015}. This could be due to the significant effect of small-scale inhomogeneities in the wind \citep[e.g.][]{Puls2008, Sundqvist2013, Puls2015} and/or to the outdated solar mixture used in the \cite{Vink2001} models, namely $Z_\odot \simeq 0.02$ with an \cite{Anders1989} mixture.  The more recent solar composition with $Z_\odot \simeq 0.014$ of \citet{Asplund2009} could reduce the discrepancy between predicted mass-loss rates and observations  (see Section~\ref{sec:metal} for a short discussion of the effects of metallicity on mass loss).  Nevertheless, the \cite{Vink2001} models are still widely used in stellar evolution codes, and we also used their recipe to calibrate $k$  to qualitatively predict the impact of radiation-driven winds on the rotational evolution of massive stars.

We thus assumed $\dot{M}= \dot{M}_{\rm Vink}$ in the non-rotating case, and we calibrated $k$ with non-rotating 1D ESTER models, that is, using mass, luminosity, and effective temperature outputs from 1D ESTER models of various masses as inputs to the \cite{Vink2001} mass loss prescription, taking $v_{\rm th}$ as the thermal velocity of Fe ions, namely $v_{\rm th}\equiv (2k_{\rm B}T_{\rm eff}/m_{\rm Fe})^{1/2}$. 
Because we calibrated $k$ using the \cite{Vink2001} mass-loss prescription, each line was considered with its appropriate $v_{\rm th}$.

From now on, the calibrated $k$ is written $k'$. 
%Note that $\dot{m}$ is independent of $v_{\rm th}$.  Indeed, our calibration imposes $k' \propto v^{\alpha'}_{\rm th}$ and (38) gives $\dot{m} \propto k^{1/\alpha'} v^{-1}_{\rm th}$.
We find that $k'$ slightly varies along the main sequence and therefore had to decide which evolution state to use for calibration. We chose to calibrate our ESTER models at ZAMS. Fitting $k'$ finally gives us the following semi-empirical function $k'(T_{\rm eff})$ at $Z=0.02$, defined on both sides of the effective temperature jump,

\begin{equation}\label{k}
\begin{aligned}
&k'(T_{\rm eff}) \simeq \\
&\begin{cases}
    \exp(-2.15 \times 10^{-4}  T_{\rm{eff}} + 2.41), & \text{if $T_{\text{eff}} \leq 20~{\rm kK}$,}\\
    -3.00 \times 10^{-6}  T_{\rm{eff}} + 0.22, & \text{if $20~{\rm kK} <T_{\text{eff}} \leq T_{\text{eff}}^{\rm jump}$,}\\
        1.16 \times 10^{-6} T_{\rm{eff}} + 0.08, & \text{if $T_{\text{eff}} > T_{\text{eff}}^{\rm jump}$,}\\
  \end{cases}
  \end{aligned}
\end{equation}
where \cite{Vink2001} defined

\begin{equation}
T_{\text{eff}}^{\text{jump}} = 61.2+ 2.59\log\moy{\rho}
,\end{equation}
with $\moy{\rho}$ the characteristic wind density at 50~\% of the terminal velocity of the wind, given by 

\begin{equation}
\log\moy{\rho} = -14.94  + 3.2 \Gamma_e \ .
\end{equation}

\begin{figure}[t!]
\psfrag{x}{$Q$}
\includegraphics[width=0.5\textwidth]{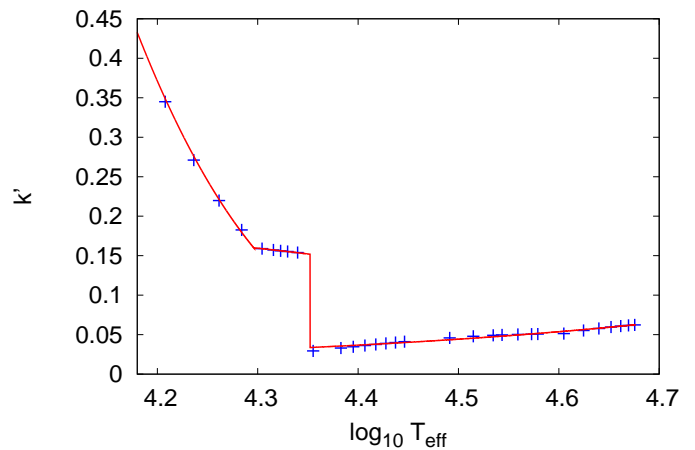}
  \caption{Calibrated FMP $k'$ as a function of the effective temperature $T_{\rm eff}$ for various 1D ZAMS models computed with ESTER for different masses with $Z=0.02$. The red full line  shows the corresponding fit.}
  \label{fig:k}
\end{figure}

The function $k'(T_{\rm eff})$ is shown in Fig.~\ref{fig:k}. Our values of $k'$ assume $v_{\rm th}=(2k_{\rm B}T_{\rm eff}/m_{\rm Fe})^{1/2}$; other assumptions on $v_{\rm th}$ would lead to other values of $k'$ to remain compatible with the \cite{Vink2001} mass loss.

\section{Effects of rotation on mass and angular momentum loss}\label{sec:lat}

After parametrising the FMPs and expressing the local mass-flux as a function of the radiative flux as well as gravity and acceleration from free-electron scattering in the non-rotating regime, we assumed that the latter follows the same scaling laws when rotation is taken into account. We therefore ignored the changes in finite disc prefactor and set this correction to $4/9$.
With Eq. (\ref{eq:1}), the local mass-flux per unit surface for a rotating star reads

\begin{equation}\label{eq:m1}
\begin{aligned}
\dot{m}(\theta) =& \frac{4}{9} \frac{\alpha(\theta) k'(\theta)^{1/\alpha'(\theta)}}{ v_{\rm th}(\theta) c} \\
 &\times\left[ \frac{c}{\kappa_e (1-\alpha(\theta))}\left(|g_{\rm eff}(\theta)|- \frac{\kappa_e F(\theta)}{c}\right)\right]^{\frac{\alpha'(\theta)-1}{\alpha'(\theta)}} \\
&\times  F(\theta)^{1/\alpha'(\theta)}\ .
\end{aligned}
\end{equation}
This local mass-flux expression is now $\theta$ dependent and thus leads to an anisotropic stellar wind that, at first glance, would favour polar ejection due to the higher polar radiative flux\footnote{Our approach implicitly assumes the presence of a weak, polewards directed component of the radiation force. Such a
non-radial component is the result of the decreasing radial velocity towards the equator, and is essential for inhibiting a flow that otherwise would be directed towards the equator. Within our approach, however, this component can be neglected when estimating the theta-dependence of $\dot{m}$. For details, see \citet{Owocki1998}, for example}. \citep[e.g.][]{Owocki1997,Owocki1998, Petrenz2000,MaederMeynet2000}. We note that \cite{Cure2004} took a different approach and analytically derived an equation for the mass-loss rate that accounts for rotation at the equator. This equation has a $\Omega$--slow solution for rotators close to criticality. We now investigate the surface distribution of $\dot{m}$ from the outputs of ESTER 2D models at ZAMS with $Z=0.02$ and for various $\omega$.%\LEt{a single sentence does not constitute a paragraph. Please either add to this or merge}.

\subsection{Latitudinal variations in mass and angular momentum loss}

We computed the local mass-flux $\dot{m}(\theta)$ as well as the local angular momentum flux 
\begin{equation}
\dot{\ell}(\theta)=\dot{m}(\theta) \Omega(\theta)R^2(\theta)\sin^2\theta \ ,
\end{equation}
with ESTER 2D models and prescription (\ref{eq:m1}).

\begin{figure}[h!]
\psfrag{y}{$\dot{\ell} \  \mbox{\small $ [M_\odot \cdot yr^{-2}]$ } $ }
\psfrag{x}{$\dot{m} \ \mbox{\small $ [g \cdot cm^{-2} \cdot yr^{-1}]$ }$ }

\includegraphics[width=0.5\textwidth]{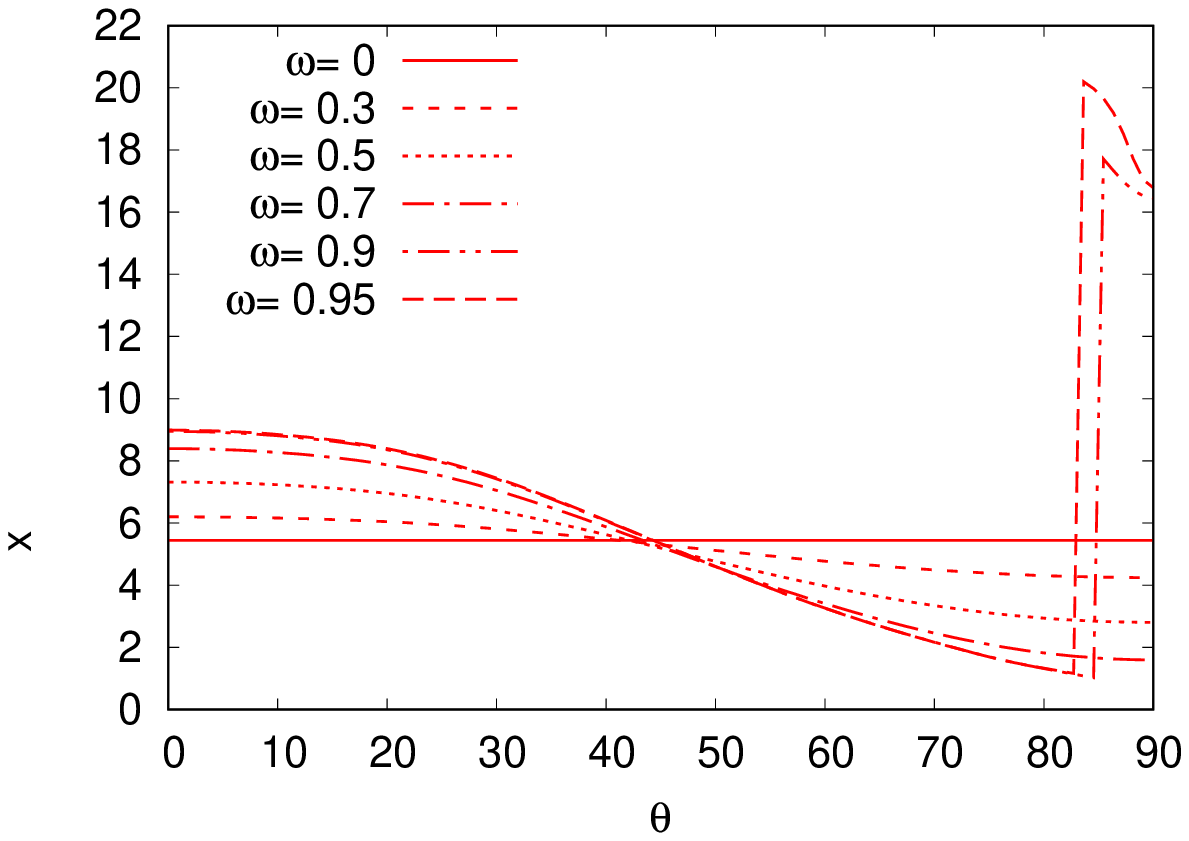} \hfill
\includegraphics[width=0.5\textwidth]{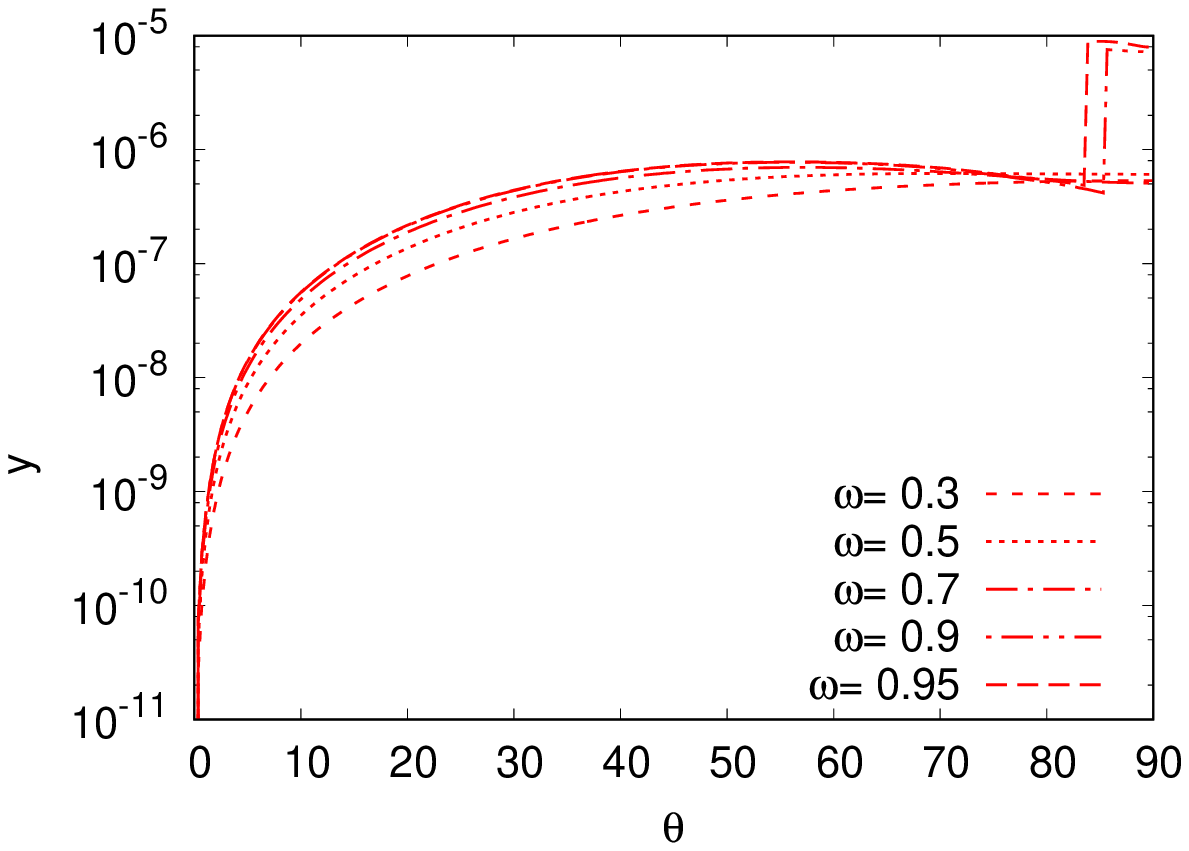}
  \caption{Variation in surface mass flux $\dot{m}$ (top) and surface angular momentum flux $\dot{\ell}$ (bottom) as a function of colatitude $\theta$  for a $15~M_\odot$ ESTER 2D-model at ZAMS with $Z = 0.02$ and various angular velocity ratios $\omega$. The main bi-stability limit is reached near the equator for $\omega \gtrsim 0.85$, and induces a strong mass-flux and angular momentum flux for colatitudes in between $\theta_{\rm jump}$ and the equator.}
  \label{fig:mdot}
\end{figure}

%Fig.~\ref{fig:mdot} shows the local mass-flux $\dot{m}$ and local angular momentum flux $\dot{\ell}$ as a function of colatitude for ESTER 2D-models of a $15 M_\odot$ star with $Z=0.02$, at ZAMS and for various angular velocity ratios $\omega=\Omega_{eq}/\Omega_k$. We see that when the star is not close to the $\Omega$-limit, mass loss is favoured in polar regions and thus decreases towards the equator. However, for sufficiently large angular velocity ratio ($\omega \gtrsim 0.85$ in Fig.~\ref{fig:mdot}), there is a colatitude $\theta_{\rm jump}$ where $T_{\text{eff}}(\theta_{\rm jump})=T_{\text{eff}}^{\text{jump}}$ and the rotation-induced bi-stability limit is crossed. In that case, mass-flux is enhanced between $\theta_{\rm jump}$ and the equator and the star is in a two-winds regime (TWR), otherwise  it is in a single-wind regime (SWR). \corr{The idea of enhanced equatorial mass-flux due to both gravity darkening and local bi-stability limit \cite{Owocki1997,Owocki1998b, Pelupessy2000} } The \corr{presence} of this rotation-induced bi-stability jump therefore strongly modifies the distribution of mass-flux with colatitude: while mass loss is favoured in polar regions in the SWR, it is a lot stronger in equatorial regions in the TWR. Similarly, while the angular momentum flux is maximum at some intermediate colatitude in the SWR, it is strongly favoured in the equatorial regions in the TWR.
Fig.~\ref{fig:mdot} shows the local mass-flux $\dot{m}$ and local angular momentum flux $\dot{\ell}$ as a function of colatitude for ESTER 2D models of a $15~M_\odot$ star with $Z=0.02$, at ZAMS and for various angular velocity ratios $\omega=\Omega_{eq}/\Omega_k$. When the star is far from the $\Omega$-limit, mass loss is favoured in polar regions and thus decreases towards the equator. However, for a sufficiently high angular velocity ratio ($\omega \gtrsim 0.85$ in Fig.~\ref{fig:mdot}), there is a colatitude $\theta_{\rm jump}$ where $T_{\text{eff}}(\theta_{\rm jump})=T_{\text{eff}}^{\text{jump}}$ and the local bi-stability limit is crossed. In that case, the mass flux is enhanced between $\theta_{\rm jump}$ and the equator and the star is in a two-wind regime (TWR), otherwise  it is in a single-wind regime (SWR). This local bi-stability jump therefore strongly modifies the distribution of the mass flux with colatitude: while mass loss is favoured in polar regions in the SWR, it is far stronger in equatorial regions in the TWR. Similarly, while the angular momentum flux is maximum at some intermediate colatitude in the SWR, it is strongly favoured in the equatorial regions in the TWR. The idea of an enhanced equatorial mass-flux that is due to both gravity darkening and the local bi-stability limit has also been discussed in the past, for instance, by \cite{Zickgraf1986}, \cite{Zickgraf1989}, \cite{Lamers1991}, \cite{Owocki1997}, \cite{Owocki1998b}, and \cite{Pelupessy2000}.

This change in the latitudinal distribution of the angular momentum flux is particularly important for stellar evolution. In the SWR, polar-dominated mass loss allows rapidly rotating massive stars to lose mass during the MS without losing much angular momentum, hence keeping a rapid rotation throughout their evolution. In the TWR, however, mass loss becomes highly dominated by the equatorial regions and the star loses far more angular momentum. This enhanced loss of angular momentum in the TWR could therefore induce a drop in $\omega$ during stellar evolution. This phenomenon will be discussed in  the follow-up paper and is not to be confused with the bi-stability braking introduced by \cite{Vink2010}, which is purely one-dimensional and corresponds to the global transition between the hot and cold side of the bi-stability jump.
We note that a star need not be close to Keplerian rotation to reach the local bi-stability limit. A rotating star that has an equatorial effective temperature that is only slightly higher than the temperature of the jump can reach the TWR with a small increase of $\omega$.

\subsection{Effects of rotation on the global mass and angular momentum loss rates}

We now compute the global mass and angular momentum loss rates by integrating $\dot{m}(\theta)$ and $\dot{\ell}(\theta)$ over the distorted stellar surface as follows: 

 \begin{equation}
\dot{M} = 2 \pi \int \dot{m}(\theta) R^2(\theta) \sqrt{1+\frac{R^2_{\theta}}{R^2({\theta)}}}\sin\theta d\theta \ ,
\end{equation}

 \begin{equation}
\dot{\mathcal{L}} = 2 \pi \int \dot{\ell}(\theta) R^2(\theta) \sqrt{1+\frac{R^2_{\theta}}{R^2({\theta)}}}\sin\theta d\theta  \ ,
\end{equation}
where $R(\theta)$ is the $\theta$-dependent radius of the star. The area element at the stellar surface is
 \begin{equation}
 dS=R^2(\theta)\sqrt{1+\frac{R^2_{\theta}}{R^2({\theta)}}}\sin\theta d\theta d\varphi \ ,
 \end{equation}
 where $R_{\theta}=\partial R/\partial \theta$ \citep{Rieutordal2016}.

\begin{figure}[t!]
\includegraphics[width=0.5\textwidth]{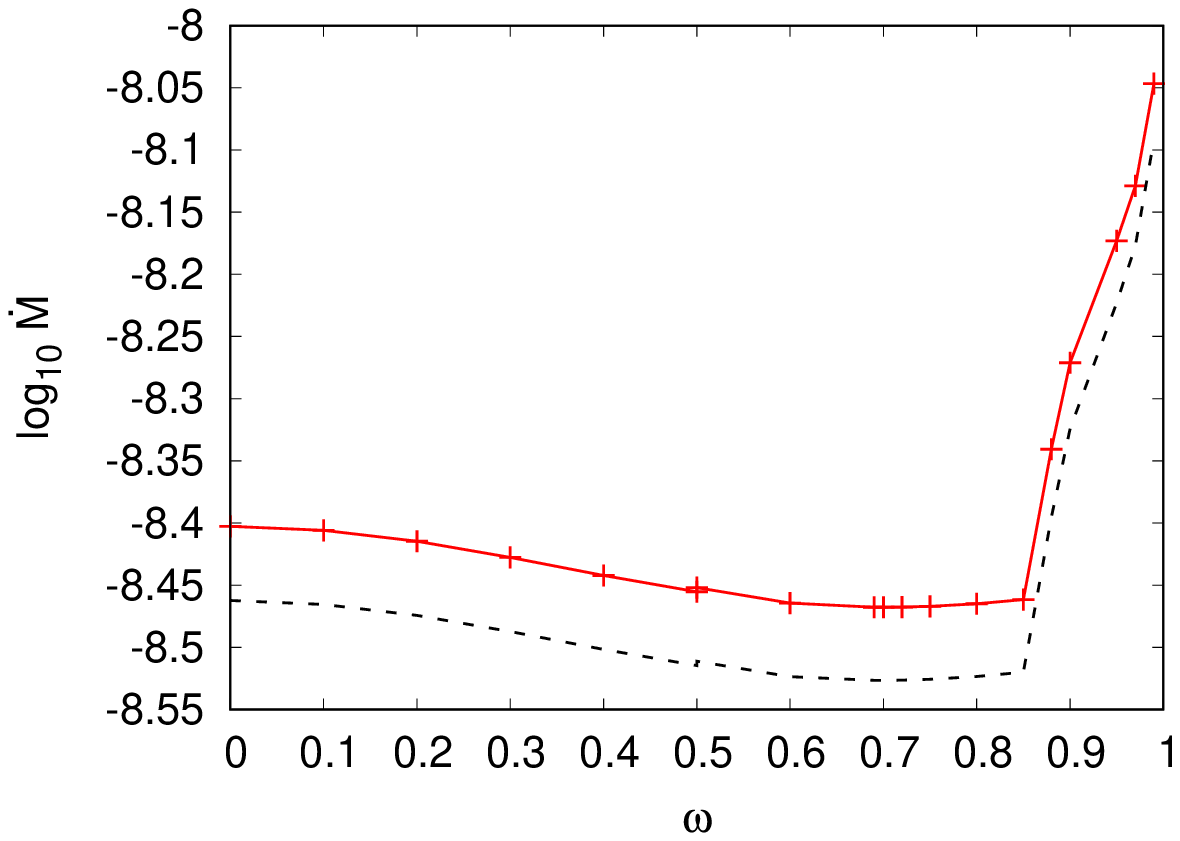} \hfill
\includegraphics[width=0.5\textwidth]{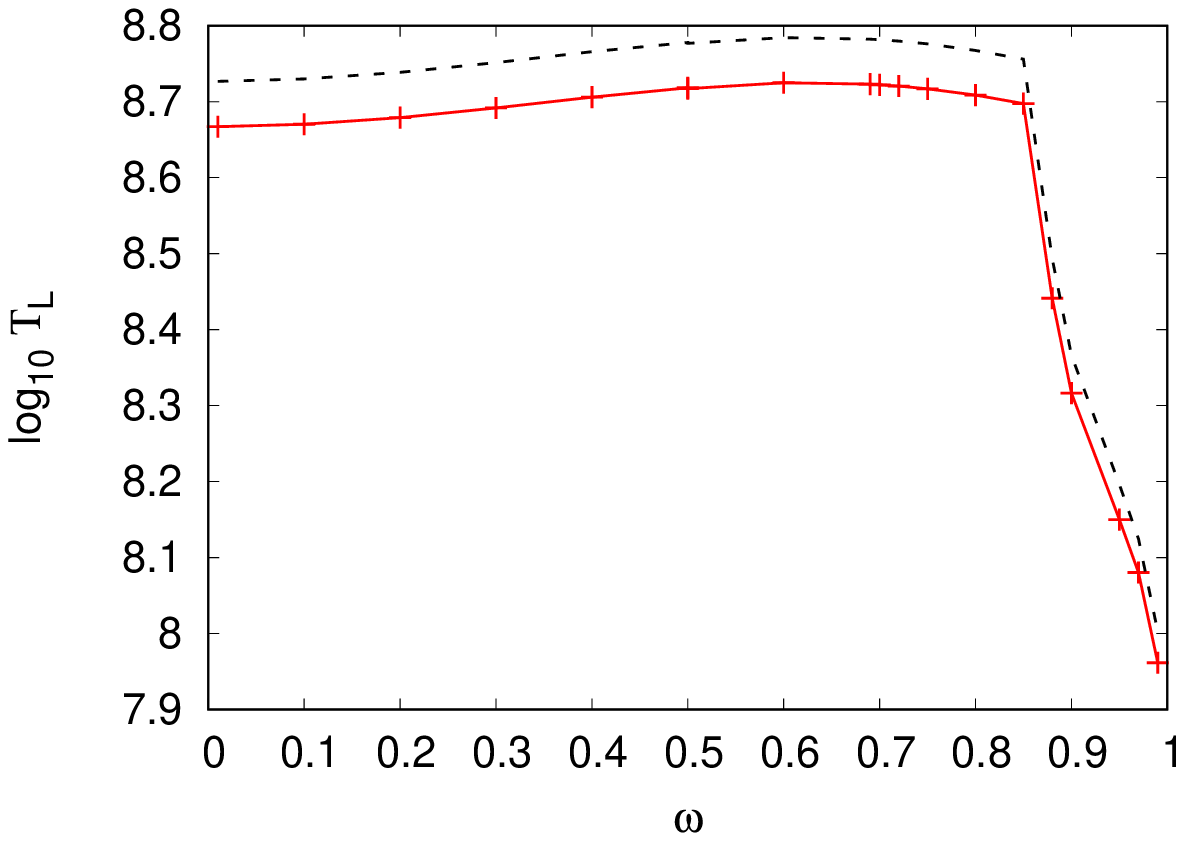}
  \caption{Variation in mass loss rate $\dot{M}$ (in $M_\odot \cdot \rm{yr}^{-1}$, top) and the angular momentum loss timescale $\Tau_\mathcal{L}$ (in yr, bottom) as a function of the angular velocity ratio $\omega$ for a $15~M_\odot$ star at ZAMS with $Z = 0.02$. The dashed lines show the same as the solid line, but with an FMP $\alpha'$ that has been reduced by 1\% to show the sensitivity of $\dot{M}$ and $\Tau_\mathcal{L}$ to FMP variations.}
  \label{fig:mdot_tau}
\end{figure}

 The global mass-loss rate $\dot{M}$, the critical angular velocity ratio $\omega_c=\Omega_{eq}/\Omega_c$ as given by the $\omega$-model (Eq.~\ref{eq:omegac}), the ratio of equatorial angular velocity to Keplerian angular velocity $\omega=\Omega_{eq}/\Omega_k$ , and the angular momentum loss timescale $\Tau_L= \mathcal{L}/\dot{\mathcal{L}}$ are reported in Table~\ref{table:1} for a $15~M_\odot$ star ESTER 2D model at ZAMS and at $Z=0.02$.
For this stellar model in the SWR, we find the global mass-loss rate to slightly decrease for increasing $\omega$, for instance, $\dot{M}(\omega=0.6)/\dot{M}(\omega=0) \simeq 0.87$ (see Fig.\ref{fig:mdot_tau} top). Similar results have been obtained by \cite{Muller2014}.

 With increasing $\omega$, the total angular momentum of the star $\mathcal{L}$ increases, and even though the global mass-loss rate $\dot{M}$ decreases in the SWR, the global loss of angular momentum $\dot{\mathcal{L}}$ also increases in this regime. This is simply because $\dot{\mathcal{L}}$ increases for increasing $\omega$.
 It is even more interesting, however, that the timescale of angular momentum loss $\Tau_\mathcal{L}=\mathcal{L}/\dot{\mathcal{L}}$ is approximately independent of the degree of criticality $\omega$ in the SWR (see Figure~\ref{fig:mdot_tau}, bottom).

On the other hand, the TWR is characterised by a strong increase in global  mass and angular momentum loss rates. In this regime, $\Tau_\mathcal{L}$ rapidly decreases as $\omega$ approaches unity.
Both the strong increase in $\dot{M}$ and decrease in $\Tau_\mathcal{L}$ can be explained by the increasing stellar surface fraction where the effective temperature  is lower than $\Tej$ as $\omega$ increases (see Fig.~\ref{fig:mdot}). 

\begin{figure}[h!]
\psfrag{x}{$\overline{T}_{\rm eff}({\rm kK})$ }
\centerline{\includegraphics[width=0.5\textwidth]{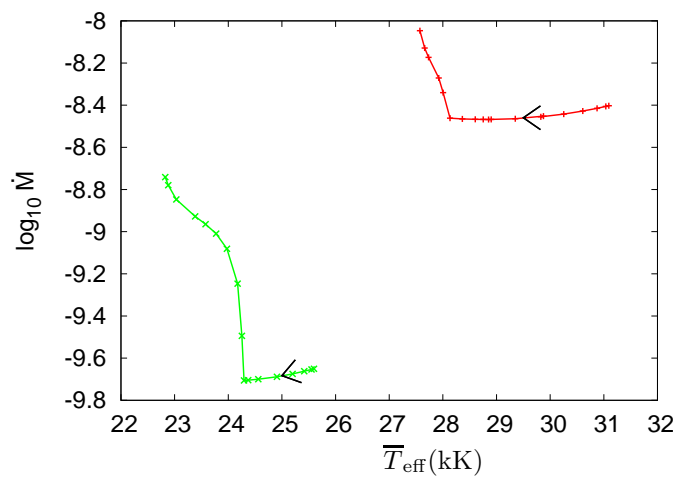}}
 \caption{Global mass-loss rate $\dot{M}$ (in $M_\odot \cdot \rm{yr}^{-1}$) for a $15~M_\odot$ (red) and a $10~M_\odot$  (green) ESTER model against the corresponding mean effective temperature $ \overline{T}_{\rm eff}$, at ZAMS with $Z = 0.02$ and for $\omega \in\interval[open right]{0}{1}$. Arrows indicate the direction of increasing $\omega$.}
  \label{fig:mdot_Teff}
\end{figure}

That $\dot{M}$ only gradually increases with increasing $\omega$ in the TWR is a result specific to 2D models. In 1D models, the bi-stability jump is accounted for with a stronger global mass-loss rate if the \textit{\textup{mean effective temperature}} of the star is lower than $\Tej\sim 22.5$--$25~{\rm kK,}$ according to \cite{Vink2001}. In the present work however, 2D models reach the bi-stability limit if the \textit{\textup{local effective temperature}}  somewhere on the stellar surface is lower than $\Tej$. This difference has two consequences. Firstly, 2D models can reach the bi-stability limit even with an average effective temperatures higher than $\Tej$. In Fig.~\ref{fig:mdot_Teff}, we illustrate the global mass-loss rate for a variety of angular velocity ratios at ZAMS for a 15~$M_\odot$ and a $10~M_\odot$ ESTER model at ZAMS and with $Z=0.02$, against the corresponding  surface-averaged effective temperature $\overline{T}_{\rm eff}$ of the model. %This is illustrated in Fig.~\ref{fig:mdot_Teff}, which shows the variation of the mass loss rate\corr{, resulting from the calculation of the local mass-flux integrated over the stellar surface,} as a function of the \corr{corresponding} surface-averaged effective temperature $\overline{T}_{\rm eff}$ for a $M= 15~M_\odot$ and a $M= 10M_\odot$ model at ZAMS with $Z = 0.02$ and for various $\omega$. 
In these models, $\overline{T}_{\rm eff}$ is greater than $\Tej\simeq 22.8~{\rm kK}$ for all $\omega$. Thus, equivalent 1D models would just have ignored the bi-stability jump.

Secondly, in 2D models the surface fraction where the effective temperature is lower than $\Tej$ monotonically increases with increasing $\omega$. This results in a gradual variation in $\dot{M}$ (and $\Tau_\mathcal{L}$) with $\overline{T}_{\rm eff}$ in the TWR (see Fig.~\ref{fig:mdot_Teff}). Hence, in rotating stars the bi-stability jump does not induce a discontinuity of the global mass-loss rate (but it induces a discontinuity of the local mass-flux, see Fig.~\ref{fig:mdot}) as the $\omega$ parameter increases (and therefore as the mean effective temperature decreases). The discontinuity occurs only on the derivative of the function $\dot{M}(\teff)$. This is further discussed in the follow-up paper.

These points show that even though the bi-stability jump might eventually be confirmed observationally (although its location in terms of mean effective temperature would be scattered, see Fig.~\ref{fig:mdot_Teff}), a full 2D spectral analysis is required to verify both qualitative and quantitative features of radiation-driven winds from rapidly rotating massive stars \citep[e.g.][]{Petrenz1996}. As a first step, a way around this full analysis would be to select stars with a small $v\sin i$ to select either slowly rotating stars or stars that are viewed pole-on. Obviously, the precise determination of $v\sin i$ for hot massive stars is a challenge in particular because rotational effects are mixed with other line-broadening effects such as the so-called macro-turbulence \citep[e.g.][]{SimonDiaz2007}.

\begin{table}[h!]
\caption{Summary of the main results for ESTER 2D models of a $15~M_\odot$ star with $Z=0.02$ at ZAMS. The first column reports the ratio of equatorial angular velocity to Keplerian angular velocity $\omega$, the second column reports the  critical angular velocity ratio $\Omega_{eq}/\Omega_c$, the third column is the global mass-loss rate $\log \dot{M,}$ and the last column gives the ratio between total angular momentum and angular momentum loss rate $\log {\rm T}_\mathcal{L}=\log \mathcal{L}/\dot{\mathcal{L}}$. $\dot{M}$ is in $M_\odot \cdot {\rm yr}^{-1}$ and ${\rm T}_\mathcal{L}$ in yr.}

\label{table:1}      
\centering    
\begin{threeparttable}                               
\begin{tabular}{c c c c}   
\\
\hline 
\\        
$\omega$&  $\Omega_{eq}/\Omega_c$  & $\log \dot{M}$ & $\log \Tau_L$\\
\hline     
 \\    
                  
    0 & 0 &$-8.40$ &  -- \\    
    \\    
    0.1 & 0.105 &  $-8.41$ &  8.60\\
    \\
    0.2 & $0.210$ &  $-8.42$ &  $8.61$\\    
    \\      
    0.3 & $0.313$ & $-8.43$ & $8.62$\\    
    \\     
    0.4 & $0.415$ & $-8.44$ &  $8.63$\\    
    \\      
    0.5 & $0.516$ &   $-8.46$ &  $8.64$\\    
    \\      
    0.6 & $0.615$ &   $-8.47$ &  $8.65$\\    
    \\      
    0.7 & $0.714$  & $-8.48$ & $8.65$\\    
    \\ 
    0.8 & $0.812$ &  $-8.49$ &  $8.64$\\
    \\
    0.9\tnote{a} & $0.909$ &  $-8.27$ & $8.35$\\    
    \\           
    0.95\tnote{a} & $0.957$ &  $-8.25$ & $8.22$\\    
    \\         
    0.99\tnote{a} & $0.995$ &  $-8.19$ & $8.13$\\    
    \\             
\hline                                             
\end{tabular}
\begin{tablenotes}
\item [a] Star in the TWR 
\end{tablenotes}
\end{threeparttable}
\end{table}

\subsection{Metallicity effect}\label{sec:metal}

Before we conclude this paper, a few words on low-metallicity stars are in order. Metallicity is indeed known to have an important effect on radiatively driven winds because metallic lines, which significantly contribute to opacity, weaken and eventually disappear. As a consequence, the FMPs, such as $\alpha$ or $k,$ are expected to decrease with a decreasing $Z$ \citep[][]{Kudritzki1987, Puls2000, Puls2008}. Moreover, mass loss is very sensitive to the value of the FMPs. In particular, a small decrease in $\alpha'$ leads to a significant decrease in $\dot{M}$ (see Figure~\ref{fig:mdot_tau}). In addition, a low metallicity causes stars to be more compact and therefore have a higher effective temperature \citep{maeder09}. This effect may compensate (partly?) for the loss of opacity on the wind acceleration. All in all, because the FMPs have a significant influence on mass-loss calculations and because they are ill-known at metallicities much lower than solar, we do not venture any prediction on the behaviour of mass flux at low Z. We leave this question to future investigations.

\section{Discussion and conclusions}\label{sec:conclu}

We investigated two questions that are a prerequisite to the study of the evolution of massive rapidly rotating stars: (i) What is the critical angular velocity of a star when radiative acceleration is significant in its atmosphere? (ii) How do the mass and angular momentum loss rates depend on the stellar rotation rate?

 To the first question, we answer that the critical angular velocity is very close to the classical Keplerian angular velocity at the equator, at least for stars with masses lower than $40~\msun$  (and for $Z=0.02$). The role of radiative acceleration turns out to be rather limited because of the combination of a reduced opacity and reduced flux at the equator. The reduction of the flux, the so-called gravity darkening, is less than was predicted by the von Zeipel model. This latter point is the main difference between this study and the pioneering investigations of \cite{Maeder1999} and \cite{MaederMeynet2000}. ESTER 2D models indeed show that the flux is almost anti-parallel to gravity in the stellar radiative envelope \citep{ELR11}. To a very good approximation, we can therefore write $\vF=f(r,\theta)\vg_{\rm eff}$, which is the base of the $\omega$-model \citep{ELR11, Rieutordal2016}. We showed that the \om\ remains close to full 2D ESTER up to rotation as high as 90\% of the critical rotation. When equatorial rotation approaches Keplerian rotation, $f(r,\theta)$ diverges at the equator, while in the von Zeipel model it remains finite. This means that the effective temperature decreases more slowly at the stellar equator than what is predicted with the von Zeipel recipe. $f(r,\theta)$ is also a monotonically increasing function of co-latitude. Its maximum is therefore reached at the equator, hence it turns out that the total acceleration $\vg_{\rm tot}=\vg_{\rm rad}+\vg_{\rm eff}$ vanishes first at the equator, when rotation is increased. Unlike the von Zeipel approximation, ESTER 2D models never predict that the radiative flux vanishes at the equator. Critical rotation, defined as the rotation required for $\vg_{\rm tot}$ to vanish somewhere at the surface, is therefore always reached before the equatorial rotation reaches Keplerian rotation. This point has been made by \cite{MaederMeynet2000}. However, 2D models hold that this difference is tiny. Considering a massive star of 40~$M_\odot$, we therefore find that criticality, $\vg_{\rm tot}=\vzero$ at the equator, is reached at $\omega \sim 0.96$ and even at 0.997 for a 15~$M_\odot$ star. This tiny difference can be understood because gravity darkening in the $\omega$-model is weaker than in the von Zeipel model and because the effect of rotation on opacity leads to a strong decrease in standard Eddington parameter towards the equator. To return to the debate  between \cite{Langer1997,Langer1998} and \cite{Glatzel1998}, our results support the latter concerning the influence of the Eddington limit on the value of critical rotation: this influence is quite small and never exceeds 4\% as far as we could test (i.e. M$\leq$ 40~$M_\odot$, $Z=0.02$).
The fact that only a small equatorial region becomes unbound at criticality may lead to mechanical mass loss. This will be discussed in a forthcoming work.
%\corr{As for the $\omega$-model itself, we show in the appendix that the estimation of the radiative flux at the surface of rapidly rotating massive stars is in good agreement with that from calculations using ESTER 2D-models. In particular, we find the relative difference between the two never to exceed 10~$\%$ for stellar models rotating at rates up to $90~\%$ of critical rotation, and to remain under 1~$\%$ when under $50~\%$ of critical angular velocity. We conclude that this analytical model is quantitatively reliable for most rapidly rotating massive stars and could be used as a replacement for radiative flux models based on the von Zeipel's law, in 1D stellar codes.}

To address the second question, we first devised a prescription for the surface density of the mass flux based on current knowledge of radiatively driven winds. The derivation of this local mass flux was based on the approaches of \cite{CAK1975} and \cite{Pauldrach1990}, but force multiplier parameters were adjusted to match the widely used prescriptions of \cite{Vink2001} for $\dot M$ in the range $\teff\in[10,50]{\rm kK}$. This prescription led to a discontinuity in the mass flux when $\teff$ drops below 22.5--25~{\rm kK}. Because the surface effective temperature of rotating stars can span a wide range of values from poles to equator, it easily happens that the discontinuity occurs at some latitude of the star. In this case, the stellar wind experiences two regimes, one centred on the poles, the other around the equator. We have shown that if the star experiences a single-wind regime (no latitude of discontinuity), the maximum extraction of angular momentum occurs at mid-latitude, while the mass flux is maximum at the poles. However, if the jump in mass flux occurs at some latitude, then both mass loss and angular momentum loss are maximum in equatorial regions. Interestingly, these two regimes are expected to affect not only the evolution of the stellar rotation rate, but also the internal rotational mixing because the applied torque is different in both intensity and location.

Before we conclude, we wish to caution about one important simplification of ESTER 2D models. The current ESTER models indeed assume that no mass flux leaves the photosphere and a zero normal velocity is imposed at the surface of the star. Moreover, we assume that as in 1D models, the surface layers are vertically in hydrostatic equilibrium. All these approximations are acceptable for determining the bulk structure of the star, but are likely too rough to describe the surface layers of a wind-emitting massive star. In particular, the values of the surface opacity, which is important for determining the radiative acceleration, may be modified when a better coupling between the wind and the star is introduced. With such a new 2D model of the wind launch region, the concept and conditions of critical angular velocity will have to be revisited. With the current models, predictions are therefore indicative: they are reliable for intermediate-mass stars (lower than $10~\msun$), but their realism and their reliability decrease with increasing mass. Beyond $40~\msun$, new models are probably mandatory to obtain a sensible description of the mass-loss phenomenon with rotation.

Finally, on the observational side, we remark that rotation makes verifying the existence of the jump in the relation $\dot{M}(\teff)$ more difficult. This verification would be possible if we could select stars whose rotation axis is aligned with the line of sight. In that case, we would be sure to face the same (polar) wind regime. If no selection can be made, the random orientation of the rotation axis means that the observed winds are sourced by an unconstrained range of $\teff$, implying that any discontinuity in the $\dot{M}(\teff)$ relation is smoothed out, unless we can reproduce the observed star with a complete 2D wind+star model. In a follow-up paper \citep{gagnier_etal19b}, we apply these results to  study the evolution of rotation in early-type fast-rotating stars and address the question, among others, how a wind can prevent a massive star from reaching the critical rotation.

\begin{acknowledgements}
We are particularly grateful to Joachim Puls for his detailed reading, comments, and suggestions on the original manuscript. We thank Georges Meynet and Fabrice Martins for enlightening discussions. We are grateful to Sylvia Ekström for providing information to validate our scheme for main sequence temporal evolution. We thank CALMIP -- the computing centre of Toulouse University (Grant 2017-P0107). M. Rieutord  acknowledges the strong support of the French Agence Nationale de la Recherche (ANR), under grant ESRR (ANR-16-CE31-0007-01), and of the International Space Science Institute (ISSI) for its support to the project ``Towards a new generation of massive star models'' lead by Cyril Georgy. F. Espinosa Lara acknowledges the financial support of the
Spanish MINECO under project ESP2017-88436-R.
\end{acknowledgements}
\bibliographystyle{aa}
\bibliography{bibnew}

\begin{appendix}

\section{Opacity dependence on effective temperature}
\label{appendixB}

In this appendix, we show that gravity darkening at the surface of rotating stars may lead to a decrease in opacity $\kappa$ towards the equator. To do this, we assumed that the opacity at the stellar surface follows Kramer's opacity law, namely,

\begin{equation}
\kappa(\theta) \propto \rho_s (\theta) T_{\rm eff}(\theta)^{-7/2}
,\end{equation}
where $\rho_s(\theta)$ is the local surface density. This approximation seems to be rather well verified at the surface of rapidly rotating ESTER 2D models.
We recall that the pressure at the surface is

\begin{equation}
P_s(\theta) = \tau_{s}\frac{g_{\rm eff}(\theta)}{\kappa(\theta)} \propto \frac{g_{\rm eff}(\theta)}{\kappa(\theta)}  \ ,
\end{equation}
where $\tau_s \simeq 2/3$ is the Rosseland mean optical depth at the photosphere. Assuming the power law $\teff\propto\geff^\beta$ and the ideal gas equation of state, the previous expression
leads to 

\begin{equation}
\kappa(\theta) \propto  T_{\rm eff}(\theta)^{-9/4 + 1/(2\beta)} \quad {\rm with} \quad \rho_s(\theta) \propto T_{\rm eff}(\theta)^{5/4 + 1/(2\beta)} .
\end{equation}
The value $\beta=0.25$ given by von Zeipel's law implies that the opacity increases towards the equator, that is, with decreasing effective temperature. When $\beta < 2/9\simeq 0.222$, however, this simple model shows that the surface opacity decreases towards the equator. Because $\beta$ decreases with rotation, $\beta < 2/9$ corresponds to a surface flattening $\epsilon \gtrsim 0.08$ (or to an angular velocity ratio $\omega \gtrsim 0.4$, according to ESTER 2D models). This scaling relation is only approximate. Still, it shows that a weaker gravity darkening than that of von Zeipel may have a strong effect on the latitudinal variations in surface density, thus on opacity at the surface of rotating stars. In other words, in some cases, rotation may induce a decrease in opacity towards the equator because of a corresponding reduced density in these regions. %\corr{Note that the opacity has a minimum that is set by pure electron scattering.}

\section{Short presentation of ESTER models}
\label{appendixD}
The ESTER code computes the steady state of an isolated rotating star,
including the large-scale flows driven by the baroclinicity of the radiative
regions. It solves in two dimensions (assuming axisymmetry) the
steady equations of stellar structure with fluid flows, namely the Poisson equation, 

\begin{equation}
\Delta\phi=4\pi G\rho \ ,
\end{equation}
 where $\phi$ is is the gravitational potential; the continuity equation,
\begin{equation}
\na\cdot\rho\vv = 0 \ ,
\end{equation}
the momentum equation,
\begin{equation}
\rho\vv\cdot\na\vv = -\na P -\rho\na\phi +\vF_{\rm visc} \ ,
\end{equation}
where $\vF_{\rm visc}$ is the viscous force; and the heat balance equation,
\begin{equation}
\rho T\vv\cdot\na s = \na\cdot(\khi\na T) + \eps_* \quad {\rm in \;radiative\;envelopes}
\end{equation}
and
\begin{equation}
\partial_r s=0 \quad {\rm in\; convective\; cores.}
\end{equation}
This last equation assumes an efficient convection in convective cores, as can be shown with the mixing-length model.

These equations are completed by boundary conditions that require that
(i) the gravitational potential $\phi$ vanishes at infinity, (ii)
velocity fields meet stress-free conditions at the stellar surface,
(iii) that the surface radiates like a local black body, and (iv) the
surface is defined by the place where the  pressure $P$ equals
$g_{\rm eff}/\kappa$. Usual notations have been used: $G$ is the gravitational constant, $\vv$ the velocity field, $s$ the entropy, and $\eps_*$ the energy produced by nuclear reactions per unit mass. 

Regarding the micro-physics, opacity and the equation of state
are given by the OPAL tables using the GN93 mixture \citep{GN93}. It might be argued that the use of the GN93 mixture is questionable considering that a newer solar chemical composition have been determined (e.g. Asplund 2009, Przybilla et al. 2008), but it is sufficient because this newer composition is not so different from the solar mixture used in \cite{Vink2001} \citep[namely][]{Anders1989}. The
diffusive transport of momentum is  ensured by a vanishingly low
viscosity, implying that no heat is advected by meridional circulation
(this is the zero Prandtl number limit). However, differential rotation
resulting from the baroclinic torque is computed as well as the associated
meridional circulation.  Nuclear energy generation is described by an
analytical formula including the pp- and CNO cycles. A more detailed description can be
found in \cite{ELR13} and \cite{Rieutordal2016}.

\end{appendix}

\end{document}